\documentclass[submission, Phys]{SciPost}
\usepackage{xcolor}
\usepackage{tablefootnote}
\usepackage{tikz}
\usetikzlibrary{
    arrows.meta,
    positioning,
    calc,
    fit,
    decorations.pathreplacing
}

\newcommand{\zbc}{z_{\mathrm{bc}}}
\newcommand{\zbcs}{z_{\mathrm{bc}}^{\,\prime}}
\newcommand{\zaux}{z_{\mathrm{aux}}}
\newcommand{\zbs}{z_{\mathrm{bs}}}
\newcommand{\zdepth}{z_{\mathrm{depth}}}

\usepackage[utf8]{inputenc} 
\usepackage[T1]{fontenc} 	
\usepackage[english]{babel} 

\usepackage[bitstream-charter]{mathdesign}
\usepackage{pifont}
\usepackage{geometry} 		
\usepackage{mathtools} 		
\usepackage{float} 			
\usepackage{graphicx} 		
\usepackage{tabularx} 		
\usepackage{booktabs} 		
\usepackage{color, xcolor} 	
\definecolor{Rcolor}{HTML}{E99595}
\definecolor{Gcolor}{HTML}{C5E0B4}
\definecolor{Bcolor}{HTML}{9DC3E6}
\definecolor{Ycolor}{HTML}{FFE699}
\definecolor{Vcolor}{HTML}{CFB4E0}
\usepackage{pdfpages} 		
\usepackage{extarrows} 		
\usepackage{multirow} 		
\usepackage{multicol} 		
\usepackage{enumitem} 		
\usepackage{xspace} 		
\usepackage{stackrel} 		
\usepackage{tikz} 			
\usetikzlibrary{arrows.meta, positioning, fit, calc}
\usepackage{braket} 		
\usepackage{bm} 			
\usepackage{tensor} 		
\usepackage{slashed} 		
\usepackage{siunitx} 		
\usepackage{lastpage} 		
\usepackage{cite} 			
\usepackage[normalem]{ulem} 
\usepackage{fontawesome} 	
\usepackage{tocloft} 		
\usepackage{titlesec} 		
\usepackage{doi} 			
\usepackage{hyperref} 		
\usepackage[most]{tcolorbox} 					
\usepackage[nameinlink, capitalize]{cleveref} 	
\usepackage[nottoc, notlot, notlof]{tocbibind} 	
\usepackage[ruled, vlined]{algorithm2e} 		
\usepackage{makecell}
\usepackage[makeroom]{cancel}
\usepackage{feynmf}
\usepackage{accents}

\makeatletter
\def\BState{\State\hskip-\ALG@thistlm}
\makeatother

\makeatletter
\@ifundefined{pdfoutput}{}{\DeclareGraphicsRule{*}{mps}{*}{}}
\makeatother

\makeatletter
\DeclareRobustCommand*{\bfseries}{%
   \not@math@alphabet\bfseries\mathbf
   \fontseries\bfdefault\selectfont
   \boldmath
}
\makeatother

\DeclareSymbolFont{usualmathcal}{OMS}{cmsy}{m}{n}
\DeclareSymbolFontAlphabet{\mathcal}{usualmathcal}

\SetArgSty{textnormal}
\SetKwComment{Comment}{{\small\#}~}{}
\SetCommentSty{mycommfont}

\setitemize{itemsep=2pt,topsep=2pt,parsep=0pt,partopsep=0pt,leftmargin=*}
\setenumerate{itemsep=0pt,topsep=2pt,parsep=0pt,partopsep=0pt,labelindent=3pt,leftmargin=*}
\usepackage{diagbox} 

\definecolor{red_cb}{HTML}{e41a1c}
\definecolor{blue_cb}{HTML}{377eb8}
\definecolor{green_cb}{HTML}{4daf4a}
\definecolor{purple_cb}{HTML}{984ea3}
\definecolor{orange_cb}{HTML}{ff7f00}

\definecolor{EmeraldGreen}{HTML}{1ea78d}
\definecolor{EnglishRed}{HTML}{b02427}
\hypersetup{
	pdftitle={Nu-Mechanistic Interpretability},
	pdfauthor={Bonnet-Guerrini et al.},
	colorlinks=true, 			
	linkcolor={red!50!black}, 	
	citecolor=EmeraldGreen, 	
	urlcolor=EmeraldGreen 	
} 

\newcommand\one{\leavevmode\hbox{\small1\normalsize\kern-.33em1}}

\newcommand{\arXiv}[2][]{%
	\ifthenelse{\equal{#1}{}}%
	{\href{http://arxiv.org/abs/#2}{arXiv:#2}}%
	{\href{http://arxiv.org/abs/#2}{arXiv:#2~[#1]}}}

\def\slashchar#1{\setbox0=\hbox{$#1$}           
   \dimen0=\wd0                                 
   \setbox1=\hbox{/} \dimen1=\wd1               
   \ifdim\dimen0>\dimen1                        
      \rlap{\hbox to \dimen0{\hfil/\hfil}}      
      #1                                        
   \else                                        
      \rlap{\hbox to \dimen1{\hfil$#1$\hfil}}   
      /                                         
   \fi}

\newcommand{\tikznode}[2]{%
\ifmmode%
\tikz[remember picture,baseline=(#1.base),inner sep=0pt] \node (#1) {$#2$};%
\else
\tikz[remember picture,baseline=(#1.base),inner sep=0pt] \node (#1) {#2};%
\fi}

\def\mathswitchr#1{\relax\ifmmode{\mathrm{#1}}\else$\mathrm{#1}$\xspace\fi}
\def\mathswitch#1{\relax\ifmmode#1\else$#1$\xspace\fi}

\newcommand{\startappendices}{%
  \appendix
  \phantomsection
  \addcontentsline{toc}{section}{Appendix}%
  \addtocontents{toc}{\protect\setcounter{tocdepth}{0}}%
}

\newcommand{\stopappendices}{%
  \addtocontents{toc}{\protect\setcounter{tocdepth}{3}}%
}
\begin{document}

\begin{center}{\Large \textbf{\color{scipostdeepblue}{Finding and using interpretable latents in a neutrino foundation model with sparse autoencoders}}}\end{center}

\begin{center}\textbf{
Rapha{\"e}l Bonnet-Guerrini\textsuperscript{1,2},
Johann Ioannou-Nikolaides\textsuperscript{3},
Inar Timiryasov\textsuperscript{3},
and
Vincenzo Piuri\textsuperscript{1}
}\end{center}

\begin{center}
{\bf 1} Computer Science Department, Universit\`a degli Studi di Milano, Milano, Italy
\\
{\bf 2} INFN Sezione di Milano, Milano, Italy\\
{\bf 3} Niels Bohr Institute, University of Copenhagen, Copenhagen, Denmark \end{center}

\section*{\color{scipostdeepblue}{Abstract}}
\textbf{
We present a first application of sparse-autoencoder-based mechanistic interpretability to particle physics. Studying a neutrino foundation model pretrained on IceCube data and fine-tuned for direction reconstruction, we identify a validated atlas of physical concepts in the model representation, using a strict validation protocol consisting of held-out tests, matched nuisance controls, and replication across independent dictionary trainings. Causal interventions show that the direction head barely draws on this atlas.  Motivated by this underused information, we train an uncertainty head on the same event-level representation to predict the model's angular reconstruction error. Unlike the direction head, it depends causally on quality and brightness features from the atlas. These results suggest that mechanistic interpretability can reveal learned latent physics encoded within a model's internal representation and help design downstream tasks that exploit it.   
}

\vspace{3pt}
\noindent\rule{\textwidth}{1pt}
\begingroup
\setlength{\cftbeforesecskip}{2pt}
\setlength{\cftbeforesubsecskip}{0pt}
\tableofcontents
\endgroup
\noindent\rule{\textwidth}{1pt}
\vspace{3pt}

\section{Introduction}
\label{sec:intro}

Modern neutrino telescopes reconstruct particle properties from sparse Cherenkov light recorded in large, irregular detector volumes \cite{Katz2011HighEnergyNA, Aartsen2013EnergyRM}. In IceCube, each event is a variable-length set of photomultiplier pulses on digital optical modules (DOMs) embedded in Antarctic ice \cite{IceCubeDAQ,Aartsen2016TheIN}. Direction reconstruction depends on the visible light pattern, but also on detector geometry, ice properties, noise and trigger conditions. As machine-learning-based reconstructions become more expressive, the learned representation that carries these effects becomes an object of physics interest in its own right.

Foundation models are a natural fit in this domain. They use a shared representation from broad detector data and couple it to task-specific heads. This can reduce task-specific training cost and improve statistical efficiency, as various downstream tasks can read the same representation differently. This trend mirrors the broader success of scaling and pretraining in machine learning \cite{scalinglawKaplan2020,FMstanford2022}, and is now emerging in particle physics through masked-particle modeling, pretraining, joint fine-tuning, and neutrino-specific foundation models \cite{Kishimoto_2023pretrainingeventclass,Golling2024MaskingSSL,Vigl_2024finetuning,PolarBERT2024,Vigl_2026scaling_laws}. Foundation models thus raise a concrete interpretability problem: which structures are encoded in the shared representation, and which of them are used by a given head?

Standard post-hoc explanations alone do not answer this problem. Saliency maps, integrated gradients, Grad-CAM-like methods, and broader explainable-AI tools assign importance scores to inputs or intermediate quantities \cite{saliencymaps_2014,IG_2017,Gradcam_2019,surveyofsurvey_2023}. While these explanations are useful, they are often local, approximate, and method-dependent, and do not provide a more general explanation about the model's behavior. In particle physics, recent work has connected learned representations to jet-substructure observables using attribution methods, probes, path patching, Shapley values, and symbolic regression \cite{Vent:2025ddm,Patel2026ExplainableAF}. 

To move from feature attribution methods to understanding the model's internal representation, we turn to mechanistic interpretability \cite{MI2020thread,2025openproblemsmechanisticinterpretability,Rai2026DissectingJT}.  
Originally developed for natural language processing (NLP), mechanistic interpretability finds a distinctive setting in scientific foundation models. Although detector pulse patterns lack the familiar semantic cues of text, detector-level observables and simulated ground truth allow proposed interpretations of internal features to be tested quantitatively. This raises two questions for neutrino reconstruction: Can mechanistic interpretability recover physical concepts within the model's internal representation? If so, which features can be named, validated, and tested for causal relevance?

We study these questions with sparse autoencoders (SAEs). An SAE learns an overcomplete dictionary of directions in activation space. A single neuron can contribute to several interpretable concepts at the same time, its representations are superposed \cite{2022toymodelssuperposition,bricken2023monosemanticity}. An SAE addresses this problem by reconstructing each activation from only a few learned directions, which aim to separate concepts that overlap in the original neuron basis \cite{Cunningham2023SparseAF,Gao2024ScalingAE,Bussmann2024BatchTopKSA}.

The main criticism of SAEs is that they offer no semantic guarantees when constructing the overcomplete dictionary. A sparse feature is not automatically interpretable, and an interpretable feature is not automatically causal. Recent studies have shown that learned features can depend on the training run, underperform linear probes on labelled concepts, or fail to act as reliable steering directions even when they appear interpretable \cite{Kantamneni2025AreSA,Paulo2025SparseAT,Kulkarni2025InterpretableAS}. To address this gap, alternatives such as end-to-end sparse dictionary learning preserve downstream behavior rather than activation structure alone \cite{Braun2024IdentifyingFI}.
We therefore treat SAE interpretability as a hypothesis that must be rigorously validated. Candidate features must survive held-out tests, matched controls, dictionary variation, and interventions.

We develop such a validation protocol and apply it to PolarBERT, a BERT-like foundation model trained on IceCube pulse sequences and fine-tuned for direction reconstruction \cite{PolarBERT2024,BERT2019}. The public Monte Carlo setting provides true direction, detector-level auxiliary flag, and per-event angular errors \cite{icecube-neutrinos-in-deep-ice}, enabling rigorous hypothesis testing. Because sparse dictionaries are stochastic objects, we test our verdicts across independent SAE training seeds, dictionary objectives, and model layers. To our knowledge, this is the first application of SAE-based mechanistic interpretability to a particle-physics foundation model.

We show that PolarBERT's frozen representations contain an atlas covering detector quality, auxiliary activity, event brightness, and geometry concepts. 

Yet, the downstream direction head uses almost none of this atlas. 
Removing or overwriting identified features, both individually and in families, leaves the predicted direction essentially unchanged. Instead, one stable axis with no simple physics name dominates the aggregated Jacobian sensitivity, although local sensitivity varies substantially between events. This suggests that the shared representation contains substantial physical information that the direction objective leaves underused.

Motivated by this underused information, we train an uncertainty head to predict the angular reconstruction error. Unlike the direction head, the uncertainty head depends causally on the bright-clean read-out and related clean-side features. At $20\%$ selection efficiency, it achieves a median angular resolution of $3.2^\circ$, compared with $20.2^\circ$ for selection using the best single detector observable. The interventions show that this head causally uses physically interpretable information in the representation.
From a physics perspective this outcome is expected and this acts as a positive control showing that using an SAE we can isolate causal physical concepts.

The paper is organized as follows. Section~\ref{sec:setup} introduces PolarBERT, the sparse dictionaries, and the validation protocol. Section~\ref{sec:readouts} maps the atlas of validated read-outs and its stability across dictionary draws. Section~\ref{sec:head-relative} shows that the causal use of this atlas depends on the downstream head. Section~\ref{sec:outlook} summarizes the implications and future directions.


\section{Setup}
\label{sec:setup}

This section defines the objects used throughout the paper. We first specify the frozen PolarBERT representation and the downstream heads that read it. We then define the sparse dictionaries trained on this representation. Finally, we describe the validation protocol used to separate readable features, validated physical read-outs, and causal features.

\subsection{PolarBERT as a controlled testbed}
\label{sec:polarbert-testbed}

PolarBERT is an eight-block BERT-like transformer pretrained on IceCube pulse sequences and fine-tuned for neutrino direction reconstruction \cite{PolarBERT2024,BERT2019}. 

During pretraining, the backbone learns through masked-DOM prediction and total-charge regression from the \texttt{CLS} state. During fine-tuning, the backbone and direction head are updated jointly. We freeze the fine-tuned direction backbone when training the uncertainty heads, so that both heads share the same representation.
Across this study, we analyze the fine-tuned backbone. We use the pretrained checkpoint alone in Section~\ref{sec:head-relative} to determine which physical information was already present before direction fine-tuning.

A variable-length sequence of detected pulses represents each event. The arrival time, measured charge, digital optical module (DOM) identifier, and a Boolean auxiliary flag describe each pulse. In the public IceCube sample, this flag identifies pulses that were not fully digitized \cite{icecube-neutrinos-in-deep-ice}. The full waveform only gets read out when the hard local coincidence (HLC) condition is met (for this at least one neighboring DOM on the same string also needs to record a signal within $1\mu s$ \cite{IceCubeKagglePaper}). Pulses that do not meet this criterion are consequently more susceptible to uncorrelated photomultiplier noise, although the flag is not itself a truth-level noise label  \cite{IceCubeDAQ}.

We summarize the auxiliary content of each event as seen by the model using two observables defined on the tokenized input window. Each event contains at most $L = 127$ pulses before tokenization, with non-auxiliary pulses retained preferentially and the remaining slots filled by a uniform random sample of the auxiliary pulses. Let $\mathcal{T}_i$ denote the resulting set of input pulses of event $i$ ($|\mathcal{T}_i| = N_{\mathrm{pulse},i} \leq L$), let $\mathcal{A}_i \subseteq \mathcal{T}_i$ denote the subset carrying the auxiliary flag, and let $q_{ip}$ be the measured charge of pulse $p$. We define
\begin{equation}
    f_{\mathrm{aux},i}
      = \frac{|\mathcal{A}_i|}{|\mathcal{T}_i|},
    \qquad
    q_{\mathrm{aux},i}
      = \frac{\sum_{p \in \mathcal{A}_i} q_{ip}}
             {\sum_{p \in \mathcal{T}_i} q_{ip}} .
\end{equation}
Here, $i$ indexes events and $p$ indexes pulses within an event. The first quantity is the fraction of input pulses marked auxiliary, while the second is the fraction of the total charge in the input window carried by those pulses. For the $87\%$ of events that fit within the window, these quantities coincide with their full-event counterparts while for truncated events, they instead characterize the input actually presented to the model. 

We use the following event classes throughout:
\begin{itemize}
    \item \textbf{Clean:}
    $f_{\mathrm{aux}}<0.05$ and $q_{\mathrm{aux}}<0.05$.

    \item \hypertarget{aux-dominated}{\textbf{Auxiliary-dominated:}}
    $f_{\mathrm{aux}}>0.90$ and $q_{\mathrm{aux}}>0.80$ ($f_{\mathrm{aux}}>0.80$ when used as the contrast class for $\zbc$).

    \item Within the clean class, we additionally define two brightness
    subsets:
    \begin{itemize}
        \item \hypertarget{bright-clean}{\textbf{Bright-clean:}}
        $Q_{\mathrm{tot}}\ge160$.

        \item \textbf{Dim-clean:}
        $Q_{\mathrm{tot}}\le122.16$.
    \end{itemize}
\end{itemize}

PolarBERT maps the pulse sequence to a contextualized pulse representation. In BERT-like models, the sequence begins with a learned \texttt{CLS} token, whose hidden state aggregates information from the pulse tokens. For event $i$, we denote the complete hidden-state sequence after transformer block $\ell$ by $H_i^{(\ell)}$, including both the \texttt{CLS} state and the pulse-token states. The final transformer-block CLS-token is
\begin{equation}
    h_i = H^{(8)}_{i,\mathrm{CLS}} \in \mathbb{R}^{256}.
    \label{eq:cls-activation}
\end{equation}
Every downstream head reads only this \texttt{CLS} representation. 
These internal activations of the frozen representation are the object investigated in this study.
The frozen direction head maps the final \texttt{CLS} state to the sphere:
\begin{equation}
    g_{\mathrm{dir}}:\mathbb{R}^{256}\to S^2,
    \qquad
    \hat{u}_i = g_{\mathrm{dir}}(h_i),
    \label{eq:direction-head}
\end{equation}
and is evaluated against the Monte Carlo truth direction $u_i$ through $\Delta\psi_i=\arccos\left(\hat{u}_i \cdot u_i\right)$. The direction head predicts the direction using only the final \texttt{CLS} state $h_i$, without direct access to the individual pulse tokens. Our interventions therefore modify the head's only input, with no alternative path through which the prediction could bypass them.
The final-block \texttt{CLS} representation defines the common vector space for sparse dictionaries, probes, principal components, and interventions. Later checks compare this final summary to information available at earlier layers. We justify the analysis of the specific layer $8$ through a layer-wise scan in Appendix~\ref{app:protocol-details}.

Crucially, our protocol trains the SAE purely on raw model activations, without requiring MC truth labels. This ensures that the methodology applies directly to real detector data, where  truth information is absent or incomplete. Here, PolarBERT serves as a testbed: the main concept criteria are defined from detector-level observables, while Monte Carlo truth provides additional controls and evaluation quantities.

The analysis uses a one-million-event sample forward-passed through PolarBERT's fine-tuned backbone and partitioned into five disjoint sets. The sets, presented in Table~\ref{tab:sae-splits}, separate sparse-autoencoder training, reconstruction validation, concept discovery, independent concept validation, and causal testing. This separation is important because it ensures that we do not use the same Monte Carlo labels to find a candidate feature, tune its controls, and evaluate the final claim.

\begin{table}[h]
\centering
\small
\begin{tabular}{llr}
\hline
Split name & Role & Fraction \\
\hline
\texttt{sae\_train} & SAE weight training & $75\%$  \\
\texttt{sae\_reconstruction\_val} & Reconstruction and fidelity validation & $10\%$ \\
\texttt{concept\_dev} & Candidate latent discovery & $7\%$  \\
\texttt{concept\_test} & Independent concept validation & $6\%$  \\
\texttt{causal\_test} & Independent causal interventions & $2\%$  \\
\hline
\end{tabular}
\caption{Disjoint event splits used for SAE training and downstream validation. Fractions refer to the nominal one-million-event partition.}
\label{tab:sae-splits}
\end{table}

Before training the SAE, we normalize the \texttt{CLS} activations using only the \texttt{sae\_train} split. Let $\mu$ and $\sigma$ denote the training-set mean and standard deviation. The normalized activation is
\begin{equation}
    \tilde{h}_i
    =
    \frac{h_i-\mu}{\sigma}.
    \label{eq:activation-standardization}
\end{equation}
We apply the same fixed transformation to all the other splits. We perform all SAE training and decoding in this normalized activation space.

\subsection{Sparse dictionaries}
\label{sec:sae-construction}
Intuitively, an SAE rewrites each \texttt{CLS} summary as a longer, low-frequency activation, so that the resulting few activations per event are more easily interpretable than the original dense vector. 
\paragraph{Architecture and sparsity.}
An SAE represents $\tilde{h}_i$ through an overcomplete sparse code $z_i \in \mathbb{R}^{m}$ with $m\gg d$, where $d=256$ is the PolarBERT activation dimension. The decoder reconstructs the normalized activation as

\begin{equation}
    \widehat{\tilde{h}}_i
    =
    b_{\mathrm{dec}} + W_{\mathrm{dec}} z_i
    =
    b_{\mathrm{dec}} + \sum_{j=1}^{m} z_{i,j} f_j,
    \label{eq:sae-decoder}
\end{equation}
where $b_{\mathrm{dec}} = D(0)$ is the learned decoder bias, i.e. the baseline reconstruction when all latent activations are zero, and $f_j=(W_{\mathrm{dec}})_{:,j}$ is the $j$-th decoder direction. Each latent coordinate $z_{i,j}$ is therefore a candidate feature activation, while the corresponding decoder vector $f_j$ defines a direction in the normalized PolarBERT activation space.

The encoder first maps the normalized \texttt{CLS} representation to a vector of latent scores, or pre-activations $a_i = W_{\mathrm{enc}}\tilde{h}_i + b_{\mathrm{enc}}.$
Then, instead of using an $L_1$ penalty to encourage sparsity, we impose a hard sparse budget with BatchTopK \cite{Bussmann2024BatchTopKSA}. For a batch of size $B$, BatchTopK keeps the largest $B_k$ positive pre-activations across the batch and sets the rest to zero:
\begin{equation}
    z_{i,j}
    =
    \max(a_{i,j},0)\,
    \mathbf{1}\left[\max(a_{i,j},0) \geq \tau_B\right],
    \label{eq:batchtopk}
\end{equation}
where $\tau_B$ is the threshold corresponding to the $B_k$-th largest positive activation in the batch. This enforces an average sparse budget of $k$ active latents per event while allowing the number of active latents to vary from event to event. We use BatchTopK because its batch-level budget allows the number of active latents to adapt across events. A concept-independent sweep selects $k=16$: the sparser $k=8$ setting reduces fidelity and dictionary utilization, whereas $k=32$ increases feature density for only a marginal reconstruction gain.

\paragraph{Training objective.}
The main reconstruction loss is the mean squared error in normalized activation space,
\begin{equation}
    \mathcal{L}_{\mathrm{rec}}= \frac{1}{B}
    \sum_{i=1}^{B}
    \left\lVert
    \tilde{h}_i-\widehat{\tilde{h}}_i
    \right\rVert_2^2 .
    \label{eq:sae-reconstruction-loss}
\end{equation}
To avoid dead latents, we add the AuxK revival loss used in scalable TopK sparse autoencoders \cite{Gao2024ScalingAE}. AuxK prevents rarely used latents from becoming permanently inactive by asking them to reconstruct the part of the input that is not explained by the main sparse code. For each event, we define the detached reconstruction residual
\begin{equation}
    r_i
    =
    \operatorname{stopgrad}\!\left(\tilde{h}_i-\widehat{\tilde{h}}_i\right).
    \label{eq:sae-aux-residual}
\end{equation}
Among latents that have not recently activated, AuxK retains the $k_{\mathrm{aux}}$ largest positive pre-activations, giving an auxiliary sparse code $z_i^{\mathrm{aux}}$. The corresponding auxiliary loss is
\begin{equation}
    \mathcal{L}_{\mathrm{aux}}
    =
    \frac{1}{B}
    \sum_{i=1}^{B}
    \left\|
        r_i - W_{\mathrm{dec}} z_i^{\mathrm{aux}}
    \right\|_2^2 ,
    \label{eq:sae-aux-loss}
\end{equation}
where the decoder bias is omitted because the auxiliary path reconstructs the residual around the main reconstruction. The full training objective is then
\begin{equation}
    \mathcal{L}_{\mathrm{SAE}}
    =
    \mathcal{L}_{\mathrm{rec}}
    +
    \lambda_{\mathrm{aux}}\mathcal{L}_{\mathrm{aux}} .
    \label{eq:sae-total-loss}
\end{equation}

AuxK is empirically important in this setting: without it, the final full-scale training run leaves $286$ dead latents, whereas nonzero auxiliary coefficients eliminate dead latents without materially changing reconstruction quality.

\paragraph{Selecting the dictionary.}
To arrive at the dictionary studied in this paper, we select the optimal hyperparameter set based on reconstruction diagnostics alone without using any downstream concept label. We scan the SAE capacity, sparsity budget, and AuxK revival parameters on \texttt{sae\_reconstruction\_val}, using the activation reconstruction quality, recovered directional fidelity, latent utilization, feature density, decoder redundancy, and dead-latent count \cite{Gao2024ScalingAE,Karvonen2025SAEBenchAC}. We summarize the selected configuration based on hyperparameter studies in Table~\ref{tab:sae-selection}.

\begin{table}[h]
\centering
\small
\renewcommand{\arraystretch}{1.2}
\begin{tabular}{p{0.28\linewidth}p{0.3\linewidth}p{0.12\linewidth}p{0.20\linewidth}}
\hline
Choice & Values tested & Selected & Basis \\
Expansion $m/d$ & $\{2,4,8\}$ & $4$ & intrinsic diagnostics \\

BatchTopK budget $k$ & $\{8,16,32\}$ & $16$ & fidelity--sparsity tradeoff \\

AuxK coefficient $\lambda_{\mathrm{aux}}$ & $\{0, 2^{-8},2^{-6},2^{-5},2^{-4},2^{-3},2^{-2}\}$ & $2^{-5}$ & zero dead latents \\

AuxK budget $k_{\mathrm{aux}}$ & $\{64,256,512\}$ & $512$ & validation diagnostics \\

Dead-step threshold $\tau_{\mathrm{dead}}$ & $\{2^8,2^9,2^{10},2^{12},2^{13}\}$ & $2^8$ & validation diagnostics \\
\hline
\end{tabular}
\caption{Concept-independent SAE selection. We select the configuration using intrinsic reconstruction and sparsity diagnostics on \texttt{sae\_reconstruction\_val} set, before any concept validation or causal analysis.}
\label{tab:sae-selection}
\end{table}

We use the selected expansion-$4$, $k=16$ dictionary for all concepts and causal analyses. This label-independent selection is essential for interpreting later sparse features as unsupervised read-outs rather than as coordinates chosen for agreement with known physics labels.

We measure the reconstruction quality  by the raw coefficient of determination,
\begin{equation}
    R^2_{\mathrm{raw}}
    =
    1 -
    \frac{\sum_i
    \left\lVert\tilde{h}_i-\widehat{\tilde{h}}_i\right\rVert_2^2}
    {\sum_i
    \left\lVert\tilde{h}_i-\bar{\tilde{h}}\right\rVert_2^2},
    \label{eq:sae-r2}
\end{equation}
where $\bar{h}$ is the validation-set mean activation in normalized space. We also evaluate whether the SAE reconstruction preserves the information used by a frozen downstream head. For each validation event, we replace the original activation by its SAE reconstruction, forward the patched representation through the frozen downstream computation, and compare the induced angular deviation to a mean-ablation baseline. 
The recovered-fidelity score is
\begin{equation}
\mathrm{fid}_{\mathrm{rec}}= 1-\frac
{\mathbb{E}_i\left[\arccos\left(g_{\mathrm{dir}}(\hat h_i)\cdot g_{\mathrm{dir}}(h_i)
\right)\right]}
{\mathbb{E}_i\left[\arccos\left(g_{\mathrm{dir}}(\bar h)\cdot g_{\mathrm{dir}}(h_i)\right)\right]}
=
1-
\frac{
\mathbb{E}_i[\Delta\psi_{\mathrm{SAE},i}]
}
{
\mathbb{E}_i[\Delta\psi_{\mathrm{mean},i}]
},
\label{eq:fidelity_rec}
\end{equation}
where $\Delta\psi_{\mathrm{SAE}}$ is the angular deviation induced by SAE reconstruction and $\Delta\psi_{\mathrm{mean}}$ is the angular deviation induced by replacing the activation with its validation-set mean. The canonical dictionary reaches $R^2_{\mathrm{raw}} \simeq 0.994$ and $\mathrm{fid}_{\mathrm{rec}} \simeq 0.989$ with no dead latents. In absolute terms, replacing $h_i$ with its SAE reconstruction changes the direction-head output by a mean angular deviation of $0.65^\circ$, whereas mean ablation changes it by $56^\circ$. These two numbers set the scale of every intervention below, and rule out a lossy reconstruction as the explanation of later nulls.

To test robustness to the dictionary draw, we train four seeds with the same architecture, splits, and normalization procedure. Across seeds, $R^2_{\mathrm{raw}} = 0.9937 \pm 0.0001$ and $\mathrm{fid}_{\mathrm{rec}} = 0.989 \pm 0.001$, with zero dead latents in each draw.
We also train per-layer dictionaries on the \texttt{CLS} state of each transformer block, to test whether some concepts reside upstream of the final summary. Finally, we train functionally oriented dictionaries whose objective preserves the frozen direction-head output. These provide a stress test of whether a dictionary optimized for downstream behavior recovers nameable physical mechanisms for direction reconstruction.

The resulting dictionaries are sparse coordinate systems for PolarBERT activations, not explanations by themselves. The validation protocol below classifies latents as candidates, validated read-outs, or causal features.

\paragraph{The reference dictionary.}
Unless stated otherwise, we report quantitative SAE results for the expansion-$4$, $k{=}16$, seed-1 dictionary selected above using intrinsic metrics alone. This is a reporting convention, not a claim that this basis is privileged. Individual latent indices are specific to one training run. Section~\ref{sec:cross-seed} therefore tests which read-outs, response patterns, and causal verdicts persist across independent seeds and dictionary architectures, while functionally trained dictionaries provide a separate stress test across training objectives. We use the reference dictionary to report concrete coordinates, but treat only the replicated structures and verdicts as conclusions about the model representation. Code and model checkpoints are available on \href{https://github.com/ioannou-nikolaides/nu-foundation-model-sae}{GitHub} and \href{https://doi.org/10.5281/zenodo.23986789}{Zenodo}, respectively.

\subsection{Validation protocol}
\label{sec:validation-protocol}

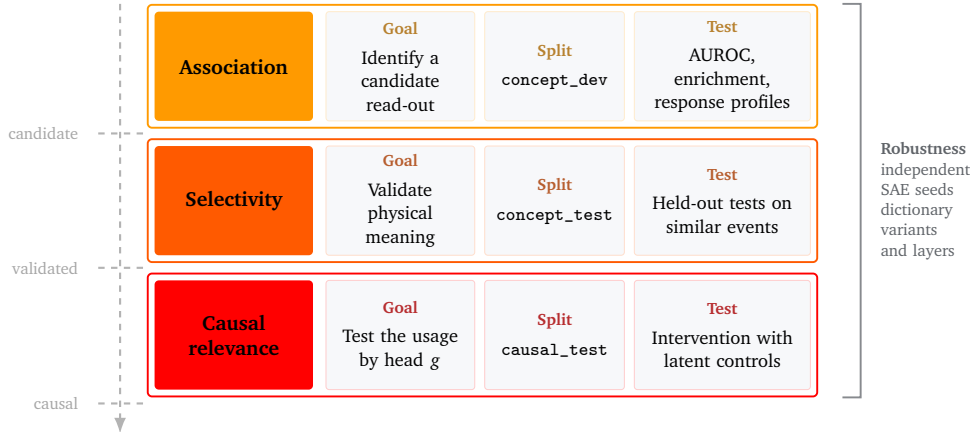
\begin{figure*}[h]
\centering

\definecolor{assocblue}{HTML}{FF9A00}
\definecolor{selectteal}{HTML}{FF5A00}
\definecolor{causalorange}{HTML}{FF0000}
\definecolor{robustgray}{HTML}{5F6368}
\definecolor{cardgray}{HTML}{F7F8FA}

\resizebox{0.88\textwidth}{!}{%
\begin{tikzpicture}[
    font=\footnotesize,
    >=Latex,
    stagehead/.style={
        font=\bfseries\small,
        rounded corners=2pt,
        align=center,
        text width=0.16\textwidth,
        minimum height=18mm,
        inner sep=3pt
    },
    subbox/.style={
        rounded corners=1.5pt,
        align=center,
        inner xsep=5pt,
        inner ysep=2pt,
        minimum height=18mm
    },
    outer/.style={
        line width=0.9pt,
        rounded corners=2pt,
        inner sep=3pt
    },
    flow/.style={
        ->,
        line width=1pt,
        gray!60
    }
]


\node[
    stagehead,
    fill=assocblue,
    text=black
] (ah) {
    Association
};

\node[
    subbox,
    draw=assocblue!20,
    fill=cardgray,
    text width=0.14\textwidth,
    right=2mm of ah
] (a1) {
    {\textcolor{assocblue!55!robustgray}{\bfseries\scriptsize Goal}}\\[1mm]
    Identify a candidate\\read-out
};

\node[
    subbox,
    draw=assocblue!20,
    fill=cardgray,
    text width=0.13\textwidth,
    right=1.5mm of a1
] (a2) {
    {\textcolor{assocblue!55!robustgray}{\bfseries\scriptsize Split}}\\[1mm]
    {\ttfamily\scriptsize concept\_dev}
};

\node[
    subbox,
    draw=assocblue!20,
    fill=cardgray,
    text width=0.17\textwidth,
    right=1.5mm of a2
] (a3) {
    {\textcolor{assocblue!55!robustgray}{\bfseries\scriptsize Test}}\\[1mm]
    AUROC,\\enrichment,\\response profiles
};

\node[
    outer,
    draw=assocblue,
    fit=(ah)(a1)(a2)(a3)
] (assoc) {};


\node[
    stagehead,
    fill=selectteal,
    text=black,
    below=4mm of ah
] (sh) {
    Selectivity
};

\node[
    subbox,
    draw=selectteal!20,
    fill=cardgray,
    text width=0.14\textwidth,
    right=2mm of sh
] (s1) {
    {\textcolor{selectteal!55!robustgray}{\bfseries\scriptsize Goal}}\\[1mm]
    Validate physical\\meaning
};

\node[
    subbox,
    draw=selectteal!20,
    fill=cardgray,
    text width=0.13\textwidth,
    right=1.5mm of s1
] (s2) {
    {\textcolor{selectteal!55!robustgray}{\bfseries\scriptsize Split}}\\[1mm]
    {\ttfamily\scriptsize concept\_test}
};

\node[
    subbox,
    draw=selectteal!20,
    fill=cardgray,
    text width=0.17\textwidth,
    right=1.5mm of s2
] (s3) {
    {\textcolor{selectteal!55!robustgray}{\bfseries\scriptsize Test}}\\[1mm]
    Held-out tests on\\similar events
};

\node[
    outer,
    draw=selectteal,
    fit=(sh)(s1)(s2)(s3)
] (select) {};


\node[
    stagehead,
    fill=causalorange,
    text=black,
    below=4mm of sh
] (ch) {
    Causal\\relevance
};

\node[
    subbox,
    draw=causalorange!20,
    fill=cardgray,
    text width=0.14\textwidth,
    right=2mm of ch
] (c1) {
    {\textcolor{causalorange!55!robustgray}{\bfseries\scriptsize Goal}}\\[1mm]
    Test the usage \\ by head $g$
};

\node[
    subbox,
    draw=causalorange!20,
    fill=cardgray,
    text width=0.13\textwidth,
    right=1.5mm of c1
] (c2) {
    {\textcolor{causalorange!55!robustgray}{\bfseries\scriptsize Split}}\\[1mm]
    {\ttfamily\scriptsize causal\_test}
};

\node[
    subbox,
    draw=causalorange!20,
    fill=cardgray,
    text width=0.17\textwidth,
    right=1.5mm of c2
] (c3) {
    {\textcolor{causalorange!55!robustgray}{\bfseries\scriptsize Test}}\\[1mm]
    Intervention with\\latent controls
};

\node[
    outer,
    draw=causalorange,
    fit=(ch)(c1)(c2)(c3)
] (causal) {};


\coordinate (flowtop) at
    ([xshift=-4.4mm]assoc.north west);
\coordinate (causallevel) at
    ([yshift=-1mm]causal.south west);
\coordinate (flowbottom) at
    ([xshift=-4.4mm,yshift=-5mm]causallevel);
\draw[flow, dashed]
    (flowtop)
    --
    (flowbottom);

\coordinate (candidatelevel) at
    ($(assoc.south west)!0.5!(select.north west)$);
\node[
    anchor=east,
    font=\scriptsize,
    text=gray!70
] at ([xshift=-10mm]candidatelevel) {candidate};
\draw[flow, dashed, -]
    ([xshift=-8mm]candidatelevel)
    --
    (candidatelevel);

\coordinate (validatedlevel) at
    ($(select.south west)!0.5!(causal.north west)$);
\node[
    anchor=east,
    font=\scriptsize,
    text=gray!70
] at ([xshift=-10mm]validatedlevel) {validated};
\draw[flow, dashed, -]
    ([xshift=-8mm]validatedlevel)
    --
    (validatedlevel);

\node[
    anchor=east,
    font=\scriptsize,
    text=gray!70
] at ([xshift=-10mm]causallevel) {causal};
\draw[flow, dashed, -]
    ([xshift=-8mm]causallevel)
    --
    (causallevel);


\coordinate (robusttop) at
    ([xshift=7mm]causal.north east |- assoc.north east);
\coordinate (robustbottom) at
    ([xshift=7mm]causal.south east);

\draw[
    draw=robustgray!75,
    line width=0.9pt
]
    ([xshift=-3mm]robusttop)
    --
    (robusttop)
    --
    (robustbottom)
    --
    ([xshift=-3mm]robustbottom);

\node[
    anchor=west,
    align=left,
    text=robustgray,
    font=\scriptsize,
    text width=0.115\textwidth
] at ([xshift=2mm]$(robusttop)!0.5!(robustbottom)$) {
    \textbf{Robustness}\\
    independent SAE seeds\\
    dictionary variants\\
    and layers
};

\end{tikzpicture}
}%

\caption{
Concept-validation protocol. A candidate SAE read-out is first identified by association, then tested for held-out selectivity, and finally intervened on to establish whether a particular downstream head uses it. Robustness checks are applied across the validation chain. Exact numerical criteria are given in Table~\ref{tab:app-thresholds}.
}
\label{fig:validation_protocol}

\end{figure*}

Our protocol validates sparse directions through three checks: association, selectivity, and causal relevance.
Throughout $z_{i,j}$ denotes the activation of latent $j$ on event $i$, and $G$ the tested object - either a single latent $G=\{j\}$ or a family of related latents. We use three terms consistently throughout the paper: the association step proposes a \textbf{candidate}, a candidate that survives the validation and matched-control tests is a \textbf{validated} read-out, and a validated read-out that additionally passes downstream intervention tests is a \textbf{causal} feature (see Figure~\ref{fig:validation_protocol}).

\paragraph{Association.}
For a concept indicated by the label $y^{(c)}$, we first rank candidate latents by their association with $y^{(c)}$ on \texttt{concept\_dev} and then re-evaluate them on an independent validation split, \texttt{concept\_test}. For binary concepts, we evaluate each latent using the AUROC of $z_j$ as a classifier and the difference in mean activation between positive and negative events.
We additionally report the firing rate, $P(z_j>0)$, and the enrichment of positive events among the highest activations
\begin{equation}
    E_q(j,c)
    =
    \frac{
    P\left(y^{(c)}=1 \mid z_j \geq Q_q(z_j)\right)
    }{
    P\left(y^{(c)}=1\right)
    },
    \label{eq:top-enrichment}
\end{equation}
where $Q_q(z_j)$ is the $q$-quantile of the latent activation. For continuous observables, we formulate the same step through response profiles or monotonic trends rather than a single binary AUROC.

\paragraph{Selectivity.}
A candidate should track the concept we claim it tracks, not a correlated nuisance variable. We therefore recompute the AUROC  after matching positive and negative events jointly on five nuisance variables: pulse count, total charge, non-auxiliary pulse count, and, in this controlled study, true zenith and azimuth\footnote{The only MC-exclusive variables used for nuisance matching during concept validation.}, and retain the candidate only if this stricter AUROC on the matched sample still clears the validation threshold. We also compare each candidate to latents with similar firing rates, excluding dense latents whose removal degrades reconstruction broadly, rather than for reasons specific to the tested concept. 

\paragraph{Causal relevance.}
We test causal relevance only after a feature passes the checks above, and always relative to a specific downstream head $g$. The simplest intervention is removal. We set the selected SAE latents to zero, reconstruct the activation, and pass it through $g$,
\begin{equation}
    \tilde{h}^{(-G)}_i
    =
    D\left(z_i \odot \mathbf{1}_{j\notin G}\right).
    \label{eq:latent-removal}
\end{equation}
Removal asks whether $g$ needs the feature. We also test whether transferring a feature's activation between contrasting event classes (e.g., from clean to auxiliary events) changes the head's prediction. In this case, we replace the activations in $G$ by donor values before decoding,
\begin{equation}
    z^{(\mathrm{write})}_{i,j}
    =
    \begin{cases}
    w_{i,j}, & j\in G,\\
    z_{i,j}, & j\notin G,
    \end{cases}
    \label{eq:latent-writing}
\end{equation}
where $w_{i,j}$ is the written activation. Writing asks whether the decoded feature can steer the head in the expected direction. We also study dose response, in which we rescale its activation, $z_j \rightarrow s\,z_j$, and measure the resulting change in the downstream head.

For each intervention we compare the effect against matched controls. In the case of removal, we use as controls latents or latent sets with a similar firing-rate distribution. Writing controls are latents with matching firing rates and the same donor activation. We fix every threshold before evaluation and report it in Appendix~\ref{app:protocol-details}. We call a feature causal for $g$ only if its effect is large enough, exceeds the matched-control effects, and is specific to the target concept. The fixed discovery, validation, control-matching, and causal-verdict rules are summarized in
Appendix~\ref{app:fixed-criteria}.

\paragraph{Supervised baseline.} Alongside the sparse latents, we use linear probes as the supervised reference: a probe is a linear model fit on the discovery split to predict $y^{(c)}$ from the full activation $\tilde{h}_i$ \cite{Alain2016:linear-probe}. Its held-out performance measures how much of the concept we can extract linearly from the full representation. 
This is the baseline against which we can judge localized sparse read-outs throughout the paper. Between these two extremes, $k$-sparse probes restricted to the $k$ most informative latents (selected on the discovery split, refit and scored on held-out events) measure how many sparse coordinates a concept actually requires.

\section{From candidates to a validated atlas}
\label{sec:readouts}
 
This section reports what the SAE-dictionary reads out of the final-layer \texttt{CLS} representation (we use probes on all layers to estimate which carries most of the directional information in Appendix~\ref{app:layer-selection}).  Of the $1024$ latents in the sparse dictionary, most do not correspond to any identifiable concepts we tested, and several concept candidates, such as event morphology and sub-detector geometry, do not correspond to any validated read-out. This is expected: when examining $1024$ latents, some may appear strongly associated with a concept purely by chance. Association alone is therefore not sufficient to validate a feature. Selectivity tests instead eliminate candidates whose apparent association is explained by other variables. The few latents that pass these tests form our validated atlas of interpretable concepts. We first present these validated read-outs and the structure they reveal in the sample, compare them against two linear baselines, and show that none of it depends on the particular dictionary. All latent indices refer to the reference dictionary of Section~\ref{sec:sae-construction} and their independence from this choice is established in Section~\ref{sec:cross-seed}.

\begin{figure}[h]
\centering
\includegraphics[width=0.98\textwidth]{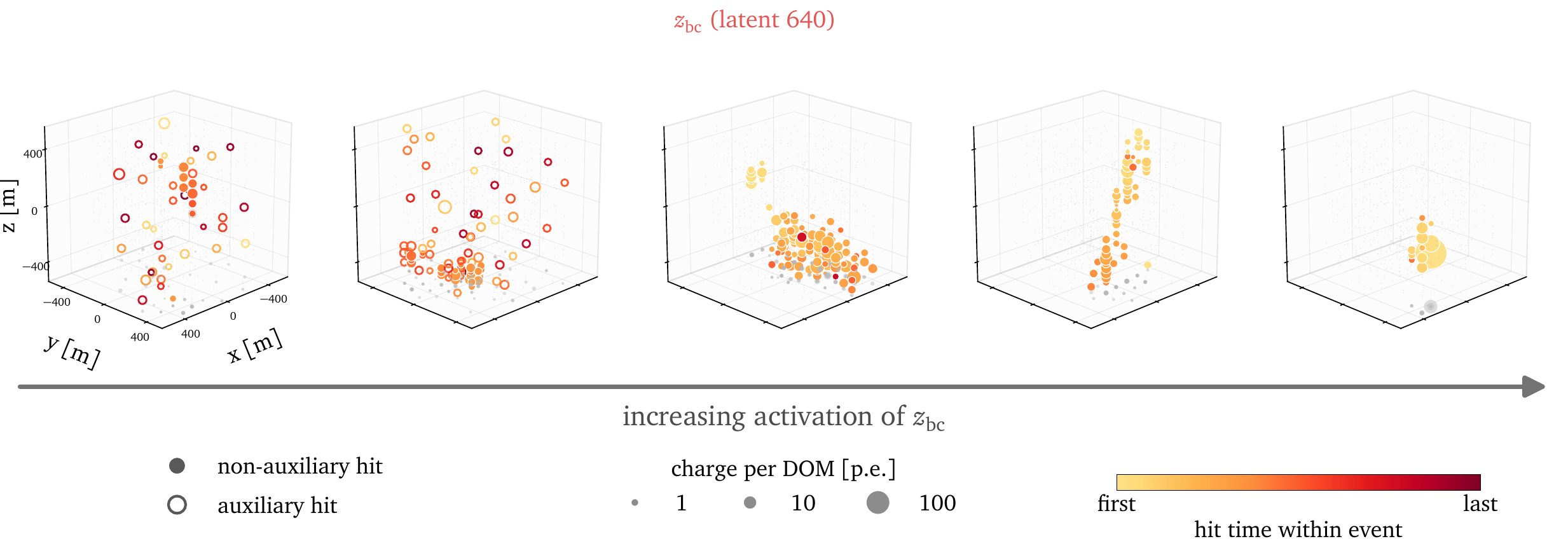}
\caption{Illustration of five events ordered by increasing activation of $\zbc$ on \texttt{causal\_test} ($z = 0.35$, $0.45$, $0.58$, $1.03$, $2.67$). Panels share camera, detector frame and charge scale.}   
\label{fig:ladder}
\end{figure}
\subsection{The identified physical atlas}
\label{sec:quality-charge-readouts}

\paragraph{Validated read-outs.}
The strongest final-layer read-out, $\zbc$, is a \hyperlink{bright-clean}{bright-clean} feature (illustrated in Figure~\ref{fig:ladder}): it activates on events whose standard, non-auxiliary pulses dominate light and whose total charge is high. In the reference dictionary this is latent $640$. On the independent validation split, it separates bright-clean events from strongly auxiliary, dim events with AUROC $0.911$, and still distinguishes the two classes when comparing events with similar brightness, pulse counts, and other properties.

\begin{table}[h]
\centering
\small
\begin{tabular}{lllll}
\hline
Feature & Role & \begin{tabular}[b]{@{}l@{}}AUROC /\\ Charge trend\end{tabular} & \begin{tabular}[b]{@{}l@{}}AUROC after\\ matching\end{tabular} & \begin{tabular}[b]{@{}l@{}}Response\\ analog\end{tabular} \\
\hline 
$\zbc$ & bright-clean events & $0.91$ & $0.79$ & $4/4$ \\ $\zbcs$  & clean events, secondary          & $0.69$& $0.84$ & $4/4$ \\ 
$\zaux$& auxiliary activity& $0.84$ & $0.85$ & $3/4$ \\ 
$\zbs$ & brightness-associated support (dense)& charge monot.\ $+0.92$& non-selective & $4/4$ \\ 
$\zdepth$ & detector depth (layer 2)& $0.986$& $0.984$ & layer bank \\
\hline
\end{tabular}
\caption{Named sparse directions identified in the paper. The third column reports the AUROC, except for $\zbs$, where it measures how strongly activation increases with charge.
The fourth column gives the lowest AUROC after matching events on potential confounding variables.
The final column indicates whether a similar response appears across dictionary seeds, rather than replication of all validation criteria (see Section~\ref{sec:cross-seed}).}
\label{tab:atlas}
\end{table} 

The complete clean- and auxiliary-side discovery rankings, including the candidates that were not retained, are reported in Appendix~\ref{app:readout-isolation}.

A weaker clean-side feature, $\zbcs$, corresponds to latent $1006$ and passes the same controls at lower discrimination. Latent $195$, $\zaux$, captures the complementary event class which activates on \hyperlink{aux-dominated}{auxiliary-dominated}, low-charge events. 
Finally, $\zbs$, latent $973$ in the reference dictionary, is a dense brightness-associated support feature. Its mean activation rises strongly with total charge, but it does not cleanly discriminate high- from low-charge events and fails the selectivity controls. We therefore do not treat it as a validated brightness read-out. Nevertheless, interventions on this coordinate produce a large change in the direction-head output. Section~\ref{sec:inert-atlas} shows that the direction reconstruction is strongly disrupted when this coordinate is moved away from its natural value. 

\begin{figure}[h]
\centering
\includegraphics[width=0.98\textwidth]{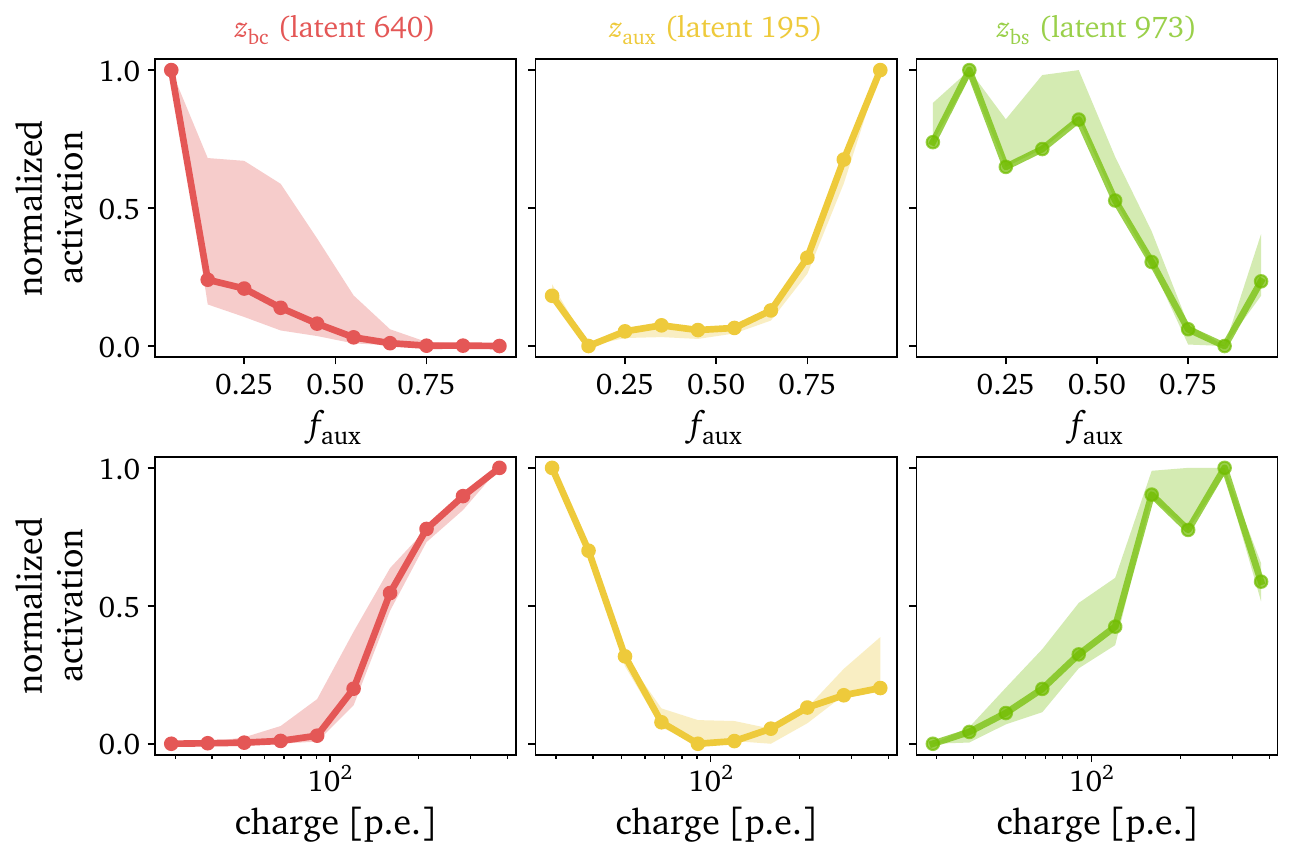}
\caption{Response profiles of the main SAE latents along the auxiliary fraction and the total charge, on  \texttt{concept\_dev} set events. Bands show the min-max spread across dictionary seeds.} 
\label{fig:profiles}
\end{figure}

\begin{figure}[h]
     \centering
     \includegraphics[width=0.98
     \linewidth]{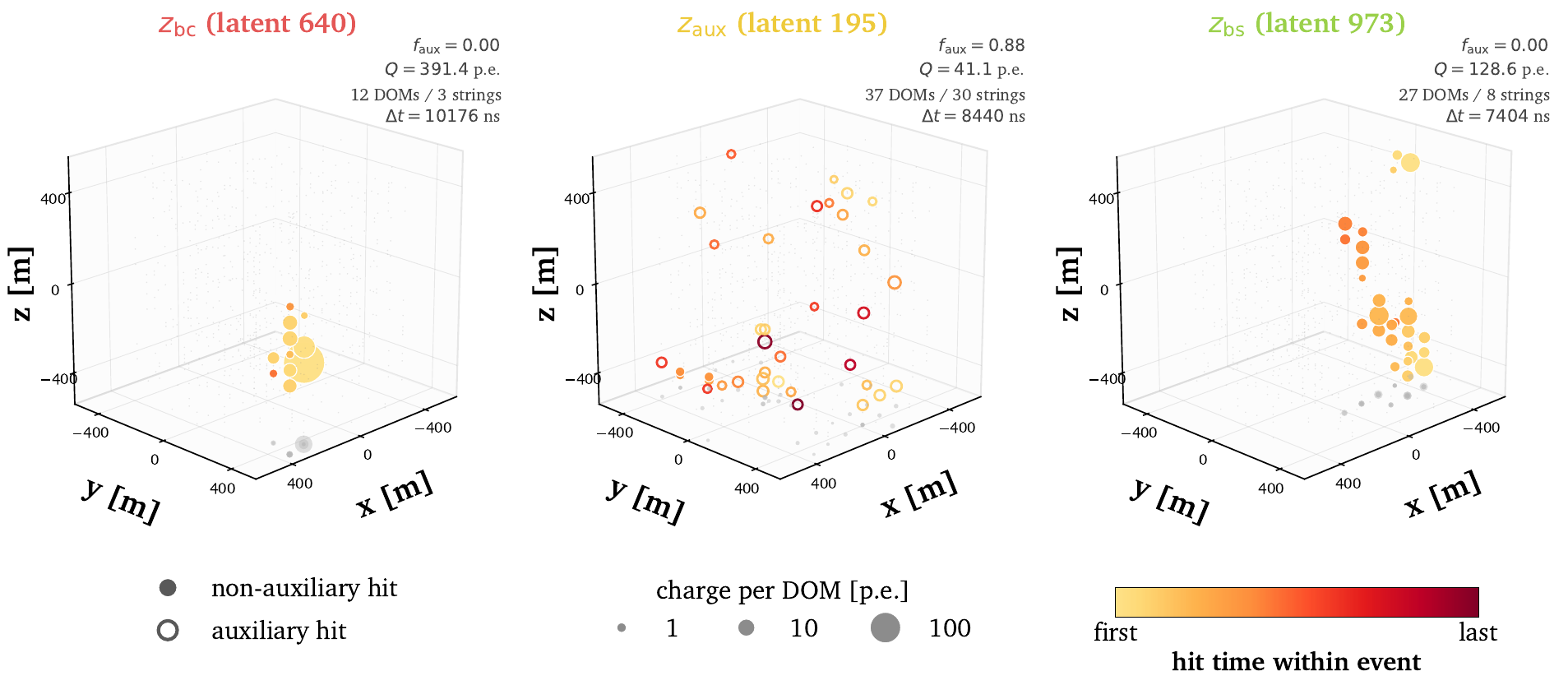}
     \caption{Highest-activating event of each latent on \texttt{causal\_test}. Markers are DOMs (charge summed per DOM), area grows with $\sqrt{\text{charge}}$, color encodes hit time within the event, and filled versus hollow marks non-auxiliary versus auxiliary hits. Panels share axes and charge scale.}
     \label{fig:3d-latent}
 \end{figure}

In Figure~\ref{fig:3d-latent}, we show the most activated event for these latents. $\zbc$ fires on a compact bright deposit ($391$ p.e., $12$ DOMs, $3$ strings). $\zaux$ fires on sparse auxiliary-dominated activity across $30$ strings. The $\zbs$ panel is not a bright event ($129$ p.e.; per-event rank correlation with total charge $0.05$ on \texttt{causal\_test}), consistent with it being a support direction rather than a concept read-out.

\paragraph{Cleanliness and brightness are one axis.}
Auxiliary-dominated events are also systematically dimmer in this sample, so auxiliary activity and brightness form a strongly correlated axis. The bright-clean and dim-clean subsets defined in Section~\ref{sec:polarbert-testbed} allow us to test brightness while holding event quality fixed. We also test the analogous brightness contrast within auxiliary-dominated events, but find no validated sparse read-out.

Within the clean class, $\zbc$ also distinguishes bright-clean from
dim-clean events, with a validation AUROC of $0.821$. Both groups satisfy the same clean-event criteria and differ only in their total-charge selection, showing that $\zbc$ also carries a brightness dependence. We therefore describe $\zbc$ as a \emph{bright-clean} read-out rather than as a pure event-quality feature. When applied to the other candidate latents, this within-clean-class test eliminates most of the weaker candidates. No latent that discriminates based on $f_{\mathrm{aux}}$ survives the analogous within-class brightness test. Thus, some response-profile shapes are real physical structures, while others arise from correlations between observables. 

\paragraph{The response landscape.}
The protocol identifies and validates a few latents, we now ask whether they are a part of a larger family. Cleanliness and brightness are broad concepts, and we find that the validated read-outs are not isolated. Instead, they sit within a broader response landscape, which we characterize by dividing all events into ten bins of auxiliary fraction $f_{\mathrm{aux}}$. For each latent $j$, we compute its mean activation in each bin and take the Spearman correlation between these ten mean activations and the corresponding $f_{\mathrm{aux}}$ bin centers. We denote this auxiliary-monotonicity score by $\mathrm{ams}_j$. Thus, $\mathrm{ams}_j\simeq-1$ indicates a latent whose activation decreases toward auxiliary-dominated events, whereas $\mathrm{ams}_j\simeq+1$ indicates one whose activation increases. We define the latent sets used later for joint interventions as:
\begin{itemize}
    \item \textbf{Clean pair:} $\{\zbc,\zbcs\}$.
    \item \textbf{Clean family:} all latents with $\mathrm{ams}\leq -0.85$, excluding the dense brightness-support direction $\zbs$; this gives $194$ latents.
    
    \item \textbf{Clean core:} the members of the clean family with firing rate $P(z_j>0)\geq0.05$; this gives $5$ latents.
        
    \item \textbf{Auxiliary family:} all latents with $\mathrm{ams}\geq +0.85$; this gives $135$ latents.
\end{itemize}
The exact reference-dictionary definitions and sizes of these intervention sets are summarized in Appendix~\ref{app:latent-families}.

A monotonic response, however, is not the same as a validated read-out. Many latents follow the global trend only weakly but do not pass every step of the validation protocol. The result is therefore not hundreds of clean or auxiliary features. It is a continuous quality-brightness axis spread across many sparse coordinates, with only a few coordinates sharp enough to validate individually. This matters for causality: if the direction head used the axis in a distributed way, removing one latent would not have a decisive impact. Section~\ref{sec:inert-atlas} therefore tests the causality of both the single validated read-outs and the full response-profile families.

\paragraph{Discoveries beyond the target concepts.}
Because we train the dictionary without labels, it can also expose structure that we do not specify in advance. Some latents vary smoothly with zenith or azimuth direction (we report them in Appendix~\ref{app:exploratory-geometry}). They are more active for events from one broad horizontal direction than from the opposite one, but no latents show the characteristic azimuthal dependence that arises from the detector's hexagonal string spacing \cite{IceCubeAzimuthalDependence}. However, given our strict protocol, none of these candidates pass the selectivity test. We still report them to illustrate the use case of sparse dictionaries for discovery: unlike a supervised probe, which requires the concept to be named before training, an SAE can surface candidates that then require thorough validation.
 
\paragraph{What the dictionary does not read, and why.}
While we find candidates for many concepts, the dictionary built from the last residual layer has an informative blind spot. DOM position information enters at the pulse-token level, so the network learns detector geometry early. At the final layer, however, no latent tied to event radius, depth, or sub-detector location survives the matched geometry contrasts. Using the same SAE recipe at layer 2 (with expansion 2) recovers a near-monosemantic depth feature, $\zdepth$, with validation AUROC $0.986$. This does not mean, however, that geometry information is lost. At earlier layers, geometry is distributed across the sequence, and later transformer blocks can still integrate information from the pulse tokens into the final \texttt{CLS} state. The early \texttt{CLS}-level depth feature is therefore not necessarily a causal bottleneck for the direction prediction. Absence of a concept from one dictionary therefore does not imply absence from the model in general. A single-layer atlas should not be read as an inventory of everything the network represents.
We similarly find no validated single-latent read-out for the elongated-versus-compact morphology contrast. We construct this contrast from charge-weighted pulse-pattern observables because the Kaggle dataset provides no track/cascade labels, so any such concept has to be identified using indirect features and would have to emerge indirectly from pulse-level structure. Linear probes nevertheless recover the continuous morphology proxies with held-out $R^2$ values between $0.46$ and $0.65$, showing that the representation contains morphology-related information without localizing it in one tested SAE coordinate.

\subsection{Sparse read-outs versus linear baselines}
\label{sec:pca-comparison}

We compare the SAE read-outs with two linear methods to better understand what the sparse dictionary adds. Principal Component Analysis (PCA) identifies directions of variation in the representation without using concept labels, while supervised linear probes test how well a concept can be predicted from the full representation.

\paragraph{Comparison with PCA} PCA identifies a similar quality-brightness structure in the tenth principal component (PC10). Unlike the SAE, which identifies separate read-outs associated with clean and auxiliary activity, PCA combines these properties into a single direction. The agreement between two unsupervised methods suggests that this structure is present in the representation itself, rather than being specific to the SAE.

\paragraph{Comparison with linear probes}

Linear probes answer a slightly different question from SAEs: they measure how well a concept is linearly decodable. A probe is trained with the concept label and can combine all 256 dimensions of the representation, which is why it is expected to perform better than an individual SAE latent. SAE latents, by contrast, are learned without these labels.
Linear probes aim to predict a concept as accurately as possible, while SAEs aim to isolate information into sparse features that can be individually identified, interpreted, and tested. For example, a linear probe predicts angular reconstruction error with AUROC 0.925, while no individual SAE latent passes our validation criteria for this task. Similarly, linear probes recover morphology-related information even though no individual SAE latent passes the morphology tests. These results illustrate both the strength and limitation of SAEs: they can provide interpretable read-outs of specific physical properties, but do not necessarily isolate all the information accessible in the full representation. Additional comparisons using sparse probes are reported in Appendix~\ref{app:morphology-probes}.

\subsection{Robustness of the atlas across dictionaries and inputs}
\label{sec:cross-seed}

We further support our finding that the foundation model's latent representation encodes a physically meaningful atlas by rerunning the same discovery, validation, and control pipeline across independent training seeds and across the different trained models of the hyperparameter studies.

Across all seeds, we find latents that have a higher activation for cleaner events. An analog of $\zbc$ appears in every seed, and in three of four seeds it satisfies the same validation criteria used for the reference dictionary. The auxiliary read-out $\zaux$ passes the validation step only in the reference seed, although auxiliary-shaped latents appear in the other seeds as well. The clean and auxiliary families also appear in every seed with similar sizes. The $\zbs$ latent rising with charge is even more stable. In every dictionary draw we recover a dense, charge-monotonic but non-selective latent. In one of the seeds, we observe that the brightness-support behavior is carried jointly by two latents rather than concentrated in one. Detailed seed-by-seed mappings and validation results are reported in Appendix~\ref{app:cross-seed}.

When re-running the clean-event isolation for $m/d \in \{2,4,8\}$ and $k \in \{8,16,32\}$, we find a control-surviving analog of $\zbc$ in every dictionary, with validation AUROC between $0.84$ and $0.99$.

As a final check, we perturb the input events themselves and re-encode them. We inject synthetic auxiliary pulses, randomly remove pulses, or rescale all pulse charges before passing the event again through the frozen backbone and SAE. The verified latents respond in the expected directions: injected auxiliary pulses raise $\zaux$ and lower $\zbc$ activation. Pulse removal suppresses $\zbc$ and $\zbs$ activation, while charge rescaling moves $\zbs$ with the charge scale. These perturbations support the physical labels of the validated read-outs and, separately, the brightness association of $\zbs$. We further test their causal role for the direction head in Section~\ref{sec:inert-atlas}.

\section{Causality is head-relative}
\label{sec:head-relative}

Section~\ref{sec:readouts} established a validated, robust atlas of concepts with a clear interpretation. However, a concept can be reliably decodable and still play no role in what a downstream head computes. This section tests that gap directly, using the removal and writing interventions of Eqs.~\eqref{eq:latent-removal} and~\eqref{eq:latent-writing}, always applied to the final-layer CLS state $h_i$. We pass every intervention through the frozen direction head and then through an uncertainty head trained on $h_i$. The two heads return opposite verdicts on the causality of the clean-side atlas. Finally, we compare the pretrained and fine-tuned backbones to test whether these physical read-outs are already present before direction fine-tuning.

\subsection{The validated read-outs are inert for the direction head}
\label{sec:inert-atlas}

\paragraph{Removal and writing on the direction head.}
As a scale reference, we replace the final \texttt{CLS} state with its validation-set mean. This intervention changes the direction-head output by $56.4^\circ$. We use this value as the baseline. 
Compared to this, zeroing $\zbc$ shifts the angular error by only $+0.034^\circ$, essentially the same as removing unrelated latents with matched firing rates ($+0.024^\circ$). This null also replicates across dictionary draws, with all analogs remaining within $|\Delta|\leq0.05^\circ$.

We repeat the removal test at every scale of the atlas. In the reference dictionary, the clean pair, clean family, clean core, and auxiliary family all give null results. The clean pair gives a null result in all four seeds. The clean family and clean core give null results in three of four seeds. In the negative seed, both family ablations change the output by about $10^\circ$ because the family includes a reconstruction-critical support latent that does not validate as a clean concept. The auxiliary family gives a null result in all four seeds.

Removing a latent is only one intervention, we also test the opposite, amplifying the activation by writing. This allows us to test whether a read-out can steer the head at all. We first write clean-like values of $\zbc$ and $\zbcs$ into auxiliary events, and conversely write an auxiliary-like value of $\zaux$ into clean events. We also repeat these interventions while removing the latent signature of the original class. In all cases, the direction prediction is unchanged. Steering brightness within the clean class is likewise null (Appendix~\ref{app:direction-writing}).
The direction head's output remains unchanged in both directions: it neither needs the quality read-outs nor can it be steered by imposing them.

\paragraph{$\zbs$ steers the direction head.}

\begin{figure}[h]
\centering
 \includegraphics[width=0.98\textwidth]{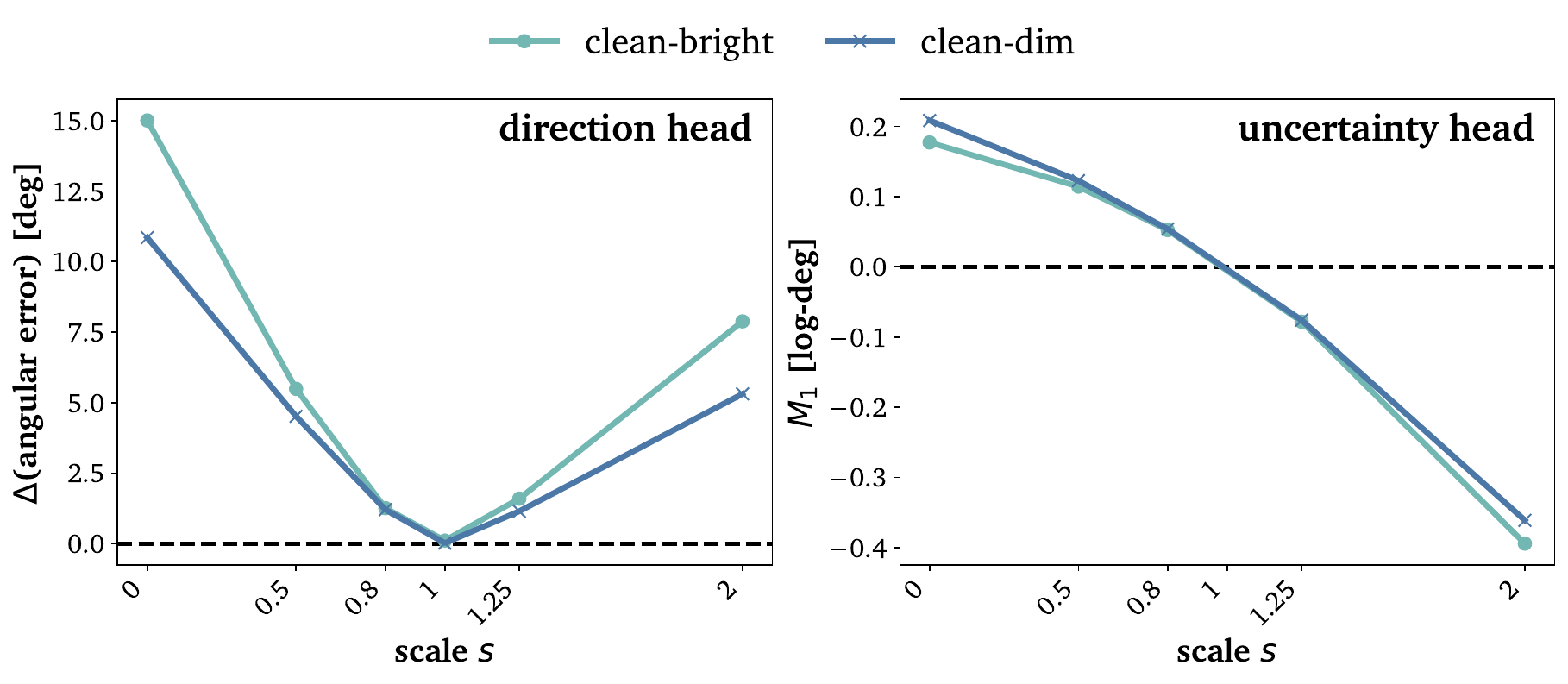}

\caption{Dose response of the brightness-support latent $\zbs$ (latent $973$) on \texttt{causal\_test}: this single coordinate is rescaled as $z\to s\,z$ before decoding. In both panels the two curves are clean events ($f_{\mathrm{aux}}<0.05$ and $q_{\mathrm{aux}}<0.05$) split by brightness alone: bright ($Q_{\mathrm{tot}}\ge160$ p.e.) and dim ($Q_{\mathrm{tot}}\le122.16$ p.e.). Left: the resulting change in the \emph{true} angular error of the direction head. Right: the resulting shift $M_1$ in the log-error \emph{predicted} by the uncertainty head (Section~\ref{sec:uncertainty-head}). }
\label{fig:dose}
\end{figure}

One dense latent, however, behaves differently from the other validated read-outs. $\zbs$, introduced in section~\ref{sec:quality-charge-readouts}, is functionally important. Setting it to zero increases the direction-head angular error by $15.0^\circ$ on clean-bright events and $10.9^\circ$ on clean-dim events. Moving $\zbs$ either below or above its natural value damages direction reconstruction systematically with a minimum angular error at its unperturbed value (Figure~\ref{fig:dose}, left). The direction head therefore does not read larger $\zbs$ as ``more brightness''. Rather, this charge-correlated coordinate supports a well-formed representation around its natural operating point. The $\zbs$-like latents recovered in all four dictionary draws show the same V-shaped impact on the angular reconstruction error.

\paragraph{Dictionary construction independence.}
Perhaps the activation dictionary is simply the wrong basis for this head. To test this, we use functionally trained (end-to-end) dictionaries \cite{Braun2024IdentifyingFI}. They are designed to preserve the output of a downstream head, pushing the recovery of causal features, which the reference dictionary misses. The learning objective of these functionally trained dictionaries is
\begin{equation}
    \mathcal{L}_{\mathrm{func}}
    =
    d\left(g_{\mathrm{dir}}(\hat h_i),\, g_{\mathrm{dir}}(\tilde h_i)\right)
    + \alpha\, \mathcal{L}_{\mathrm{rec}},
    \label{eq:e2e-loss}
\end{equation}
with $d$ an angular distance and $\alpha \geq 0$ interpolating between a local dictionary and a purely functional dictionary.

We test every causal latent against the full concept vocabulary: quality, auxiliary activity, charge, zenith, azimuth, depth, morphology, and error (Appendix~\ref{app:e2e-dictionaries}). The two purely functional dictionaries contain no bright-clean analog, with a best AUROC of about $0.53$. The two reconstruction-anchored dictionaries recover bright-clean analogs with AUROC $0.87$-$0.88$, but these latents remain outside the causal sets.

\subsection{The direction head's causal axis}
\label{sec:causal}

The previous interventions show that the direction head depends strongly on the \texttt{CLS} representation, but only weakly on the physical read-outs identified by the SAE. We therefore ask which directions in the $256$-dimensional \texttt{CLS} space the frozen head is locally sensitive to. The local sensitivity of the frozen head to the summary is its Jacobian,
\begin{equation}
    J_i = \frac{\partial \hat{u}_i}{\partial h_i} \in \mathbb{R}^{3\times 256},
    \label{eq:head-jacobian}
\end{equation}
computed exactly through the frozen head. We compute the Jacobian of the direction head for each event in the discovery split and combine these Jacobians to identify the representation directions to which the head is most sensitive. We detail the process in Appendix~\ref{app:causal-basis-construction}. 

At the dataset level we find a direction that is stable across splits. We quantify how similar the directions across different splits are by their principal angle in the $256$-dimensional space. The leading axis obtained on the \texttt{concept\_dev} differs by only $5.6^\circ$ from the axis that we compute independently on \texttt{concept\_test}. This angle lies far from the average $90^\circ$ split of two random directions. It accounts for $98\%$ of the aggregate sensitivity on the latter. This dataset-level stability does not mean every event relies on the same direction locally. For a given event, we can ask what fraction of its own local sensitivity the fixed axis captures. The typical event captures none of it - the median is 0\% because most individual Jacobians are close to degenerate to begin with. So, the head's local response is nearly flat in most directions for most events. It is the minority of events that the shared axis explains well, which is what produces the higher sample mean of 14.9\%. The stable axis is therefore a dataset-level summary of where sensitivity concentrates, not a bottleneck every event's prediction passes through.

We ask four questions about this stable axis, in the following order: does it match a named concept, is it simply the direction of largest generic variance, does the direction head depend on a physical quantity, and is it really causal rather than merely correlated with the output?

\begin{figure}[h]
\centering
\includegraphics[width=0.98\textwidth]{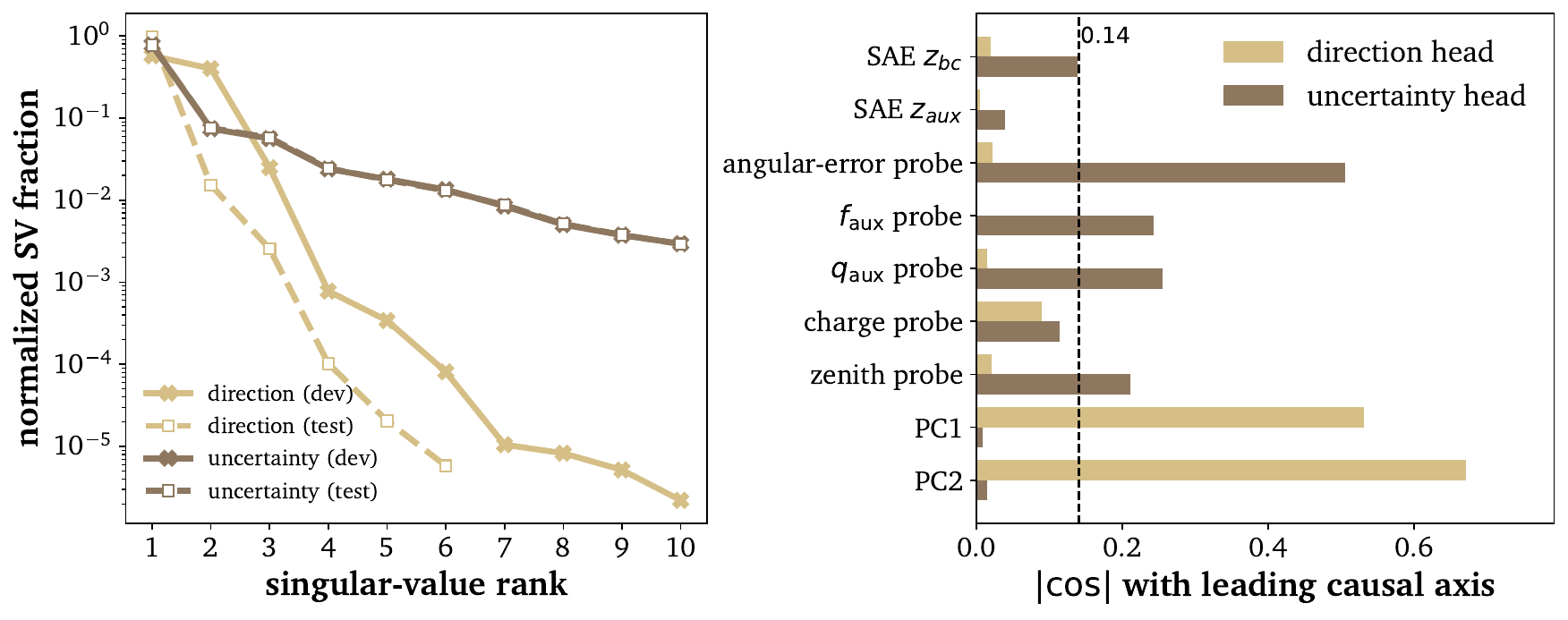}
\caption{ Sensitivity structure of the direction and uncertainty heads. Left: normalized singular-value spectra of the aggregate head Jacobians on the \texttt{concept\_dev} and \texttt{concept\_test} splits. Right: absolute cosine similarity between each head's leading sensitivity axis and selected SAE decoder directions, physics-probe directions, and the first two principal components. The dashed vertical line marks $|\cos|=0.14$. }
\label{fig:nameability}
\end{figure}

It does not match any named concept and no verified latent exceeds $|\cos|=0.15$, see Figure~\ref{fig:nameability}. However, it is also not simply a generic direction. Its strongest alignment with the two leading principal components is $|\cos|=0.62$ and $0.52$, meaning that it is related to the directions that dominate the representation's overall variance, but clearly distinct from them. 

Each probe direction in Figure~\ref{fig:nameability} is the linear-probe construction of Section~\ref{sec:validation-protocol}, which we use to test whether the Jacobian of the direction head is related to a labeled quantity. We fit each linear model to predict one labelled quantity (angular error, auxiliary fraction, auxiliary charge fraction, total charge, or zenith) from the full CLS representation. The resulting probe directions are only weakly aligned with the direction-head causal axis, with the charge probe reaching at most $|\cos|=0.1$.

The clearest evidence that the Jacobian and the direction head are causal and not merely correlated comes from the causal latents of the functionally trained dictionaries. The set of causal latents of the reconstruction-anchored functional dictionary
aligns strongly with the direction-head causal subspace, with
$|\cos|\simeq0.8$. This direction is meaningful and its misalignment with the physical latents of the named concepts is a finding, not a coincidence.

Finally, perturbing the \texttt{CLS} representation along the Jacobian-derived sensitivity directions changes the predicted direction, confirming that these directions have a causal effect. The induced change is event-dependent and does not correspond to a simple operation, such as a fixed rotation.

The direction head, then, is not indifferent to its representation, but has a real, replicable causal axis, just not one that matches any concept we can name. This raises a sharper question: is that indifference to the atlas a property of the quality and brightness features themselves, or a property of this particular head? If inertness were a property of the features, it would persist for any reader. If it is instead head-dependent, a task that actually requires event-quality information should use them, even though the direction head does not. Direction reconstruction has no explicit need to assess whether an event is clean. However, error prediction does.

\subsection{An uncertainty head with causal latents}
\label{sec:uncertainty-head}
\label{sec:same-latents}

\paragraph{Training an error reader.}
We train an uncertainty head, a one-hidden-layer MLP $U_{\mathrm{MLP}}$ ($256\to128\to1$, GELU), on the same frozen representation $\tilde{h}_i$, leaving both the backbone and direction head unchanged. It predicts
\begin{equation}
l_i
=
\log\left(1+\Delta\psi_i/\mathrm{deg}\right).
\label{eq:uncertainty-target}
\end{equation}

We also trained a simple linear regressor on the same target, which reaches comparable performance, indicating that most of the error-relevant information is already linearly accessible in the representation. All causal interventions use $U_{\mathrm{MLP}}$.

$U_{\mathrm{MLP}}$ reaches a Spearman correlation of $0.60$ with the true error on the independent validation split. 
As an additional ranking test, we use the predicted error to distinguish events in the highest and lowest quartiles of true angular error.
The MLP reaches an AUROC of $0.925$, equal to a dedicated linear classifier trained on the full $256$-dimensional \texttt{CLS} representation and well above the $0.787$ obtained from a classifier using only pulse count, total charge, $f_{\mathrm{aux}}$, $q_{\mathrm{aux}}$, and zenith.

The head recovers essentially all linearly available error information. And because that information is distributed rather than localized, the head must read broadly. It is the natural consumer of the atlas.

\paragraph{Measuring intervention effects.}
To test whether the verified concepts we discovered in $h_i$ become causal for the uncertainty head, we repeat every intervention of Section~\ref{sec:inert-atlas}. We measure the effects as the induced shift in the predicted log error,
\begin{equation}
M_1(G)
=
\mathbb{E}_i\left[
U\left(\tilde h^{(G)}_i\right) - U\left(\tilde h_i\right)
\right],
\label{eq:m1-metric}
\end{equation}
with $U$ the uncertainty head. Because the target is a logarithm, $M_1$ has a direct multiplicative reading. Independent of an event's baseline, an effect of $M_1$ corresponds to scaling $(1+\Delta\psi_{\mathrm{pred}}/\mathrm{deg})$ by $e^{M_1}$. 

We judge effects, as before, by empirical exceedance of twenty matched control interventions and a fixed effect floor. The controls are latents matched by firing rate, with dense latents that contribute strongly to activation reconstruction excluded from the control pool when appropriate.
As a scale reference, replacing $\tilde h_i$ with its training-set mean shifts the predicted log error by 0.85 on average (absolute value, all \texttt{causal\_test} events). We call a feature causal only when the intervention changes the predicted error by a meaningful amount, exceeds the effects of the matched controls, and moves the prediction in the expected direction. Appendix~\ref{app:fixed-criteria} gives the exact thresholds.

\paragraph{The clean and auxiliary latents become causal.}
As detailed in Table~\ref{tab:two-heads}, the bright-clean latent becomes causal. 
Zeroing $\zbc$ gives $M_1=+0.31$. This multiplies $\left(1+\Delta\psi_{\mathrm{pred}}/\mathrm{deg}\right)$ by approximately $e^{0.31}\simeq1.36$.
The effect is far beyond its matched controls. We repeat the causal tests across all scales of the verified atlas. The clean pair gives $M_1=+0.32$, the $194$-latent clean family gives $M_1=+0.78$, and the dense clean core gives $M_1=+0.37$. This clear causality holds across all dictionary seeds, with all effects well beyond their matched controls. 

The dense brightness-support coordinate $\zbs$ is causal for the uncertainty head as well. Unlike its V-shaped, non-monotone effect on the direction head, the uncertainty head's predicted error varies monotonically on $\zbs$ in both event classes, see Figure~\ref{fig:dose}, right (cross-seed details are in Appendix~\ref{app:uhead-crossseed}).

The auxiliary side is causal as well, but weaker. Removing the auxiliary family lowers the predicted error in all four dictionaries ($M_1$ between $-0.085$ and $-0.093$), the sign expected if auxiliary activity signals noisier events, and about nine times smaller than the clean-family effect. The single latent $\zaux$ is less robust. Its removal lowers the predicted error in every dictionary that contains an auxiliary analog, but the shift exceeds the effect threshold $|M_1|\geq0.02$ only in the reference dictionary ($M_1=-0.059$, compared with $-0.011$ and $-0.015$ in the other two), and one dictionary has no auxiliary analog at all. The auxiliary information therefore enters the uncertainty head mainly through the family rather than through $\zaux$ alone.

\begin{table}[h]
\centering
\small
\begin{tabular}{lccc|ccc}
\hline
& \multicolumn{3}{c|}{Direction head}
& \multicolumn{3}{c}{Uncertainty head} \\
Intervention
& effect & verdict & seeds
& $M_1$ & verdict & seeds \\
\hline
remove $\zbc$
& $+0.03^\circ$
& null & $4/4$
& $+0.31$
& causal & $4/4$ \\

remove $\zaux$
& $+0.00^\circ$
& null & $3/3$
& $-0.059$
& causal & $1/3$ \\

remove clean pair
& $+0.03^\circ$
& null & $4/4$
& $+0.32$
& causal & $4/4$ \\

remove clean family
& $-0.02^\circ$
& null & $3/4$
& $+0.78$
& causal & $4/4$ \\

remove clean core
& $+0.02^\circ$
& null & $3/4$
& $+0.37$
& causal & $4/4$ \\

remove aux family
& $-0.01^\circ$
& null & $4/4$
& $-0.085$
& causal & $4/4$ \\

scale $\zbs$
& V-shaped
& support & $4/4$
& monotone
& causal & $3/4$ \\
\hline
\end{tabular}
\caption{Intervention effects on the final layer for the direction head and the uncertainty head. The ``seeds'' columns report replication of the corresponding verdict across the four independently trained SAE dictionaries (three for $\zaux$, which has no analog in one dictionary). Direction-head effects are measured as shifts of angular error in degrees. Uncertainty-head effects are measured as shifts of predicted log error, $M_1$ in Eq.~\eqref{eq:m1-metric}. } 
\label{tab:two-heads}
\end{table}

\paragraph{The causal basis becomes physically interpretable.}

The intervention results tie the uncertainty head directly to named sparse read-outs.
Clean-side quality and brightness features that are inert for the direction head become causal for the uncertainty head, individually and in families, while the auxiliary family is causal with a smaller effect of the expected sign. In Figure~\ref{fig:nameability}, for the MLP uncertainty head, one aggregate sensitivity axis carries $78\%$ of the variance and is stable across the discovery and validation splits, with a principal angle of $0.4^\circ$. Unlike the direction-head axis, it also captures a median of $77\%$ of each event's own local sensitivity. This is a direction genuinely shared across events rather than an average over a mostly insensitive population. This means that the uncertainty head runs through one genuinely shared, low-dimensional mechanism across almost every event, while the direction head's computation is fundamentally distributed and event-specific with no common bottleneck.

This shared uncertainty axis is not itself a latent or one of the principal components. As expected, it aligns most strongly with the angular-error probe, followed by event-quality probes. The SAE interventions therefore establish which named features the head uses, while the probe alignment identifies the broader physical quantity the uncertainty head relies on.

Our validation protocol applied to the same final layer of the backbone gives opposite verdicts for two different heads. \textit{Represented but not used} is not a property of a feature alone, but a relation between a representation and a specific downstream head.

\paragraph{The physical read-outs predate direction fine-tuning.}
We finally compare the fine-tuned backbone with its pretrained-only checkpoint. Charge and detector geometry are already strongly recoverable before direction fine-tuning: total charge reaches $R^2=0.997$, compared with $0.960$ after fine-tuning, while depth and radius have comparable probe performance across the two checkpoints. Thus, physical structures are present and do not necessarily require fine-tuning. Instead, fine-tuning can partially degrade unused directions. 

Such an analysis can reveal which useful structures are already present and help guide the choice of what to fine-tune, which layers to unfreeze, and which information should be preserved.
Fine-tuning currently has no way to know that a concept it never explicitly needs might still be valuable to a reader trained later on the same representation, so nothing in the objective protects it. Since our protocol can track a concept's matched-control quality across checkpoints, the same procedure could in principle run during fine-tuning itself to keep concepts intact for downstream heads that do not exist yet. We do not test this here, but the degradation we observe is direct evidence that fine-tuning is not neutral with respect to information represented, which is exactly the kind of blind spot mechanistic interpretability is positioned to catch.

\subsection{Event selection and calibration}
\label{sec:selection}

The uncertainty head turns physical information underused by the direction head into a per-event estimate of angular reconstruction error. We now ask whether this estimate can identify a sample with improved effective angular resolution. For each selection variable, we orient the score $s_i$ so that larger values indicate better-reconstructed events. At a target efficiency $\epsilon$, the figure of merit is the median angular error of the retained sample,
\begin{equation}
    R(\epsilon;s) = \operatorname{median}_{\,i\in S_\epsilon(s)}
    \Delta\psi_i,
\label{eq:resolution-efficiency}
\end{equation}
evaluated on the held-out \texttt{causal\_test} split. We compare the uncertainty head with detector observables, the atlas feature $\zbc$, and distributed supervised read-outs of the same frozen representation.

\begin{figure}[h!]
  \centering
\includegraphics[width=0.98\textwidth]{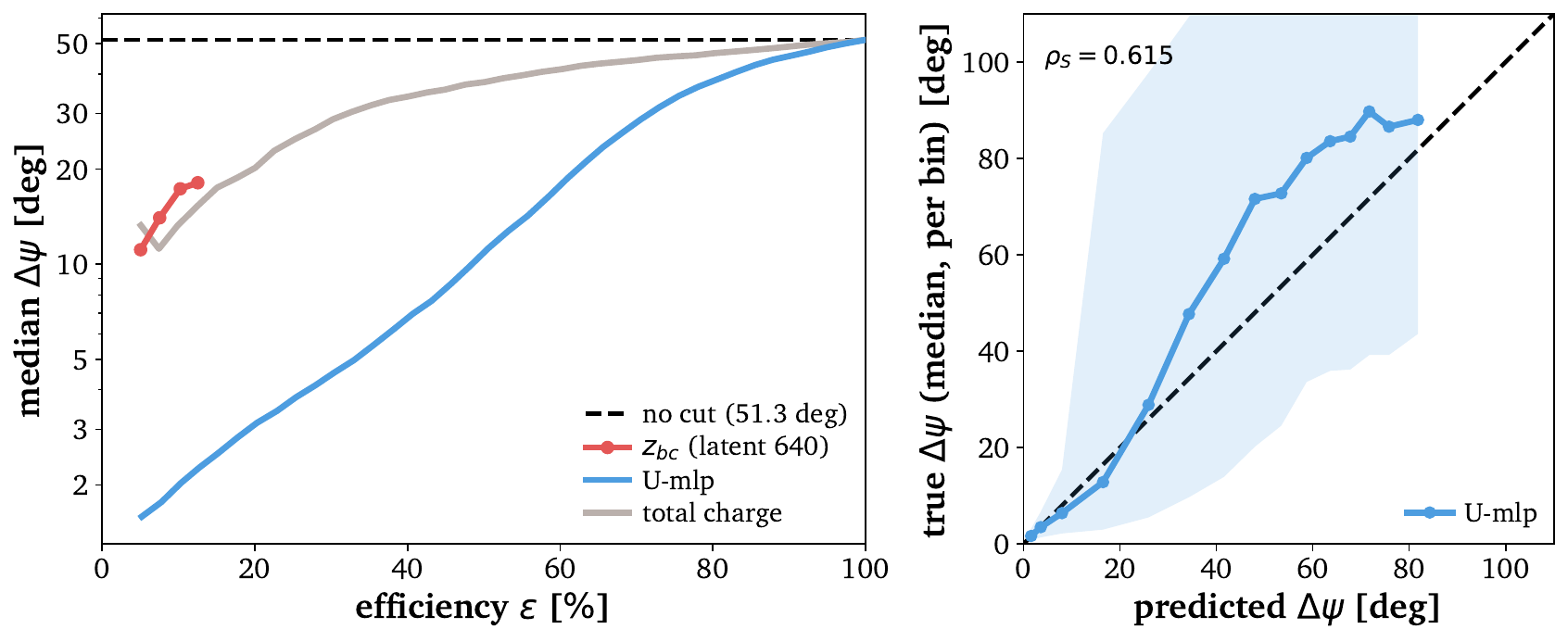}
\caption{Left: median angular error of the retained sample as a function of selection efficiency for detector-level observables, distributed representation read-outs, the sparse feature $\zbc$, and the uncertainty head. The horizontal line gives the unselected resolution, and bands show bootstrap intervals. Right: median true angular error in bins of the error predicted by $U_{\mathrm{MLP}}$. The shaded band contains the central $68\%$ of events in each bin, and the diagonal indicates perfect calibration.}
\label{fig:selection}
\end{figure}

Figure~\ref{fig:selection}-left shows the resulting resolution-efficiency curves. Without selection, the median angular error is $51.3^\circ$. Among the detector-level observables, total charge gives the strongest selection: retaining the highest-charge $20\%$ of events reduces the median to $20.2^\circ$. The representation-derived uncertainty estimates are substantially sharper. At the same efficiency, the linear and MLP heads reach $3.16^\circ$ and $3.15^\circ$, respectively. At $50\%$ efficiency they give $11.32^\circ$ and $11.24^\circ$, compared with $37.72^\circ$ for total charge. Their performance is also close to a linear probe trained on the full \texttt{CLS} representation to distinguish high- from low-error events: the probe reaches $3.41^\circ$ at $20\%$ efficiency and marginally leads at $50\%$, with $11.14^\circ$. The comparison with $\zbc$ clarifies the distinct roles of sparse interpretation and distributed prediction. Although $\zbc$ is a validated physical read-out, it is not an effective standalone error-ranking score. It fires on only $\sim14\%$ of events, so quantile thresholds targeting larger efficiencies fall on its zero-activation plateau and return the unselected sample. In this analysis, the sparse latents provide localized variables for physical validation and causal intervention, whereas accurate event ranking requires combining information distributed across the full summary.

Figure~\ref{fig:selection}-right shows the calibration curve of the uncertainty head. The true median error rises with the predicted error, with an event-level Spearman correlation of $\rho_s=0.615$. In the low-error region relevant for the tightest selections, the binned medians remain close to the diagonal, meaning that the uncertainty head is well-calibrated. At intermediate predicted errors, the uncertainty head is overconfident, and the broad within-bin spread shows substantial event-to-event variation. The head is therefore well ordered and approximately calibrated in the most-certain regime, but its raw outputs should not be interpreted as globally calibrated event-by-event uncertainties without an additional calibration step.

Together, these results show that the uncertainty head is both accurate and physically grounded. It ranks events nearly as well as a supervised probe built directly for the task, and Subsection~\ref{sec:uncertainty-head} showed that this ranking causally depends on named, validated features rather than on opaque structure.

\section{Outlook}
\label{sec:outlook}

To our knowledge, we apply sparse-autoencoder-based mechanistic interpretability to a foundation model in particle physics for the first time. We ask what physical information is represented internally and which parts are used by downstream heads. The two questions have different answers. Across its layers, PolarBERT contains validated read-outs of event quality, auxiliary activity, brightness, and detector depth. The direction head shows marginal causal dependence on the validated physical read-outs across the dictionaries we trained, including functionally trained dictionaries. Its aggregate sensitivity is instead dominated by a stable causal axis with no simple physical interpretation. What the model demonstrably represents and what the direction task uses are almost disjoint.

After this diagnosis, we investigate whether a more complex task could use this atlas. 
An uncertainty head, trained on the same event-level representation to predict the model's own error, reads the atlas that the direction head did not use. Intuitively, direction-reconstruction uncertainty should depend on how bright and clean an event is \cite{IceCubeNormFlows,IceCubeLowEReco}. Our results confirm this expectation and demonstrate that SAEs can find physically-grounded causal latents.

The SAE latents representing how clean and bright an event is are causally inert for the direction head and become causal for the uncertainty head across all four independently trained dictionaries, and the resulting estimator is a sharp selector, not just an interpretable one. At $20\%$ selection efficiency, the best single detector observable reaches the $20^\circ$ scale, while the uncertainty reader reaches $3.2^\circ$.

A black-box regressor trained on the same target might reach similar numbers, but its inputs would be opaque. Because we isolate identifiable concepts, validate them with matched controls, and test them through interventions, we propose that the uncertainty head relies on physically interpretable information that can help justify its event selection rather than requiring it to be taken on faith. This is particularly important when working with real data, where the correctness of a selection cannot be checked against Monte Carlo truth.

We thus show that \textit{represented but not used} is a relation between a shared representation and a particular downstream task. This is especially relevant for foundation models, where multiple objectives can reuse the same representations. Sparse dictionaries can reveal which physical information is available in that shared representation and which parts remain unused by a given head. This information can then guide the design of new downstream tasks or learning objectives that explicitly exploit the available structure. In this sense, mechanistic interpretability is not only a diagnostic tool, but also a way to inform how foundation-model representations can be used.

We test our protocol on Monte Carlo data, but each step of the validation protocol can also be applied to real detector data, where truth labels are incomplete and the atlas must stand on controls and replication alone. Natural next steps are to carry the uncertainty score into full likelihood analyses, repeat the interpretability study with models trained on more event-level labels such as morphology, and extend it to other particle-physics foundation models.



\section*{Code and data availability}
\label{sec:code}
The analysis code is available at \url{https://github.com/ioannou-nikolaides/nu-foundation-model-sae} and builds on PolarBERT (\url{https://github.com/timinar/PolarBERT}). The pretrained and fine-tuned PolarBERT checkpoints, the four final-layer dictionaries, the layer-2 and layer-3 dictionaries, one functionally trained dictionary, the uncertainty head, and cached inputs are archived on Zenodo (\href{https://doi.org/10.5281/zenodo.23986789}{doi:10.5281/zenodo.23986789}). The remaining dictionaries are retrained with the released code. The simulated IceCube events are taken from the public Kaggle sample~\cite{icecube-neutrinos-in-deep-ice}.

\section*{Acknowledgements}
We thank Antonin Vacheret for suggesting the potential of applying mechanistic interpretability to neutrino foundation models, Kevin Urquia for discussions and feedback on the application of sparse autoencoders, and Jean-Loup Tastet for guidance on the use of PolarBERT.
Part of this work was supported by the European Union's Horizon Europe research and innovation programme under the Marie Skłodowska-Curie grant agreement No. 101168829, Challenging AI with Challenges from Physics: How to solve fundamental problems in Physics by AI and vice versa (AIPHY). Part of this work was also supported by a research grant (VIL57416) from VILLUM FONDEN.
We thank the CINECA Leonardo HPC cluster (via INFN grant PML4HEP) and the University of Copenhagen's SCIENCE AI Centre for providing computing resources.

\bibliography{ML_ref}

\startappendices
\clearpage
\section{Supplementary analysis protocol}
\label{app:protocol-details}

This section justifies the representational site, states the fixed verdict rules, and defines the head-aligned sensitivity basis. The following sections then report additional atlas results, robustness checks, and detailed interventions in the same order as the main text.

\subsection{Choice of representational site}
\label{app:layer-selection}

To justify analyzing the final-block \texttt{CLS} state, we fit ridge probes ($\alpha=1$) \cite{Alain2016:linear-probe} from the \texttt{CLS} residual state at every depth to the true neutrino direction. Index $0$ denotes the post-embedding state, while indices $1$--$8$ denote the transformer-block outputs. We train the probes on \texttt{concept\_dev} (69{,}632 events) and evaluate them on the disjoint \texttt{concept\_test} split (59{,}648 events).

\begin{figure}[h]
\centering
\includegraphics[width=0.58\textwidth]{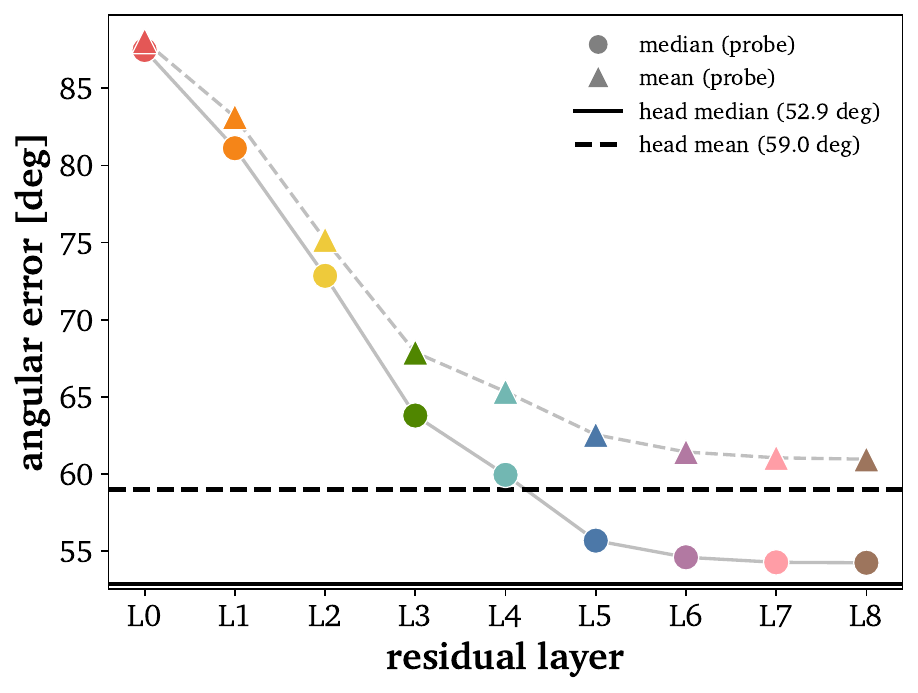}
\caption{Held-out mean angular error of a linear \texttt{CLS} probe at each layer. Lower values are better; the frozen direction head is shown as a reference.}
\label{fig:layer-scan}
\end{figure}

Figure~\ref{fig:layer-scan} shows that linearly decodable directional information improves monotonically with depth and plateaus from block~6 onward. The final-block probe reaches a mean angular error of $61.0^\circ$, close to the $59.0^\circ$ obtained by the frozen nonlinear direction head on the same events. This supports using the final-block \texttt{CLS} state as the common site for the dictionary, probes, and head interventions.

\subsection{Fixed read-out and causal criteria}
\label{app:fixed-criteria}

We fix each verdict rule before its corresponding held-out evaluation and retain partial or failed replications. For a target intervention statistic $T$, its control-normalized score is
\begin{equation}
    z(T)
    =
    \frac{T-\operatorname{mean}(T_{\mathrm{ctrl}})}
         {\operatorname{std}(T_{\mathrm{ctrl}})},
\end{equation}
where $T_{\mathrm{ctrl}}$ is obtained from the matched-control interventions. For direction-head tests, the selectivity statistic is $\Delta_{+}-\Delta_{-}$, with the control statistic computed in the same way. Table~\ref{tab:app-thresholds} collects the rules used throughout the analysis.

\begin{table}[h]
\centering
\small
\begin{tabular}{p{0.34\linewidth} p{0.58\linewidth}}
\hline
Stage & Fixed rule \\
\hline
Primary binary candidate & Discovery AUROC $\geq0.75$ and survival of every evaluable matched-control scheme \\
Secondary read-out & Reported separately when it survives the nuisance controls but remains below the primary AUROC bar \\
Independent confirmation & Re-evaluate the fixed candidate on \texttt{concept\_test}, without reselection or threshold tuning \\
Family membership & Auxiliary-monotonicity score $\mathrm{ams}\leq-0.85$ for the clean side or $\mathrm{ams}\geq+0.85$ for the auxiliary side \\
Direction-head causality & $\Delta_{+}\geq0.05^\circ$, $z(\Delta_{+})\geq2$, and $z(\Delta_{+}-\Delta_{-})\geq2$ \\
Uncertainty-head causality & $|M_1|\geq0.02$ and $|z(M_1)|\geq2$ in the pre-specified direction (two-sided when no sign is pre-specified) \\
Removal controls & $20$ firing-rate-matched latents or sets; multi-latent controls also match firing-rate composition \\
Writing controls & The same donor values as the target write, with firing-rate-matched recipient coordinates \\
Density exclusion & Control-pool exclusion for firing rate $>0.30$ or top-decile mean nonzero activation \\
\hline
\end{tabular}
\caption{Fixed discovery, validation, and intervention rules. Each row gives the condition used at that stage.}
\label{tab:app-thresholds}
\end{table}

The density rule applies to matched-control pools, not automatically to the physical families. For the within-clean brightness analysis, the control pool additionally excludes the $43$ charge-tracking latents defined by that analysis. Latent $1006$ is retained as a secondary clean-side read-out because it survives the nuisance controls despite falling below the primary discovery AUROC.

\subsection{Head-aligned sensitivity basis}
\label{app:causal-basis-construction}

For event $i$, let $v\in\mathbb{R}^{256}$ be a unit direction in the standardized \texttt{CLS} activation space. A small perturbation $\epsilon v$ changes a frozen head to first order as
\begin{equation}
    g(\tilde h_i+\epsilon v)-g(\tilde h_i)
    \simeq
    \epsilon J_i v,
\end{equation}
where $J_i$ is the head Jacobian. Thus, $\lVert J_i v\rVert$ measures local sensitivity along $v$. We compute the Jacobians by automatic differentiation, verify them against finite differences, and aggregate them over $n=8000$ discovery-split events as
\begin{equation}
    M
    =
    \frac{1}{n}\sum_{i=1}^{n}J_i^{\top}J_i.
    \label{eq:aggregate-sensitivity}
\end{equation}
All Jacobians are expressed in the standardized \texttt{CLS} frame used by the SAE, probes, and PCA.

Because
\begin{equation}
    \operatorname{tr}M
    =
    \frac{1}{n}\sum_i\lVert J_i\rVert_F^2,
\end{equation}
events for which the head is more locally responsive contribute more strongly without an explicit event weight. The eigenvectors of $M$ order activation-space directions by their average effect on the head output, and the normalized eigenvalues give their share of aggregate sensitivity. We retain a direction only when it replicates on the independent split.

\section{Supplementary atlas results}
\label{app:suppl-atlas-result}

\subsection{Isolation of the named read-outs}
\label{app:readout-isolation}

Table~\ref{tab:app-readout-isolation} reports the leading clean-event coordinates on the discovery split. Latent $640$ is the only primary candidate. Latent $1006$ remains below the primary AUROC bar but is retained as a secondary read-out because it survives the nuisance controls and concentrates clean events in its activation tail. The sparse candidates $751$, $946$, and $369$ fail the controls. Latent $973$ has the largest mean-activation contrast but is dense and non-selective, motivating its separate treatment as a support coordinate rather than a validated concept read-out.

The auxiliary contrast yields one primary candidate, latent $195$. Its discovery AUROC is $0.832$, and its held-out AUROC is $0.840$. The evaluable nuisance controls survive, with a minimum AUROC of $0.854$. Schemes that additionally match the non-auxiliary pulse count produce no usable pairs because that count is nearly deterministic for the extreme clean and auxiliary classes. The remaining candidates have either weak discrimination or high enrichment at negligible firing rates and are not promoted.

\begin{table}[h]
\centering
\small
\begin{tabular}{r c c c c l}
\hline
Latent & AUROC & Top-$1\%$ enrich. & Firing rate & Min. ctrl AUROC & Status \\
\hline
\multicolumn{6}{l}{\textit{Positive class: clean events}} \\
$\mathbf{640}$ & $\mathbf{0.913}$ & $15.40\times$ & $0.139$ & $0.818$ & primary \\
$1006$ & $0.698$ & $14.94\times$ & $0.087$ & $0.699$ & secondary \\
$751$ & $0.631$ & $6.13\times$ & $0.134$ & $0.550$ & rejected \\
$946$ & $0.601$ & $15.18\times$ & $0.014$ & $0.510$ & rejected \\
$418$ & $0.591$ & $3.89\times$ & $0.126$ & $0.590$ & rejected \\
$369$ & $0.589$ & $6.51\times$ & $0.081$ & $0.543$ & rejected \\
$973$ & $0.561$ & $2.07\times$ & $0.406$ & $0.561$ & support only \\
\hline
\multicolumn{6}{l}{\textit{Positive class: auxiliary-dominated events}} \\
$\mathbf{195}$ & $\mathbf{0.832}$ & $7.07\times$ & $0.493$ & $0.827$ & primary \\
$261$ & $0.661$ & $19.01\times$ & $0.024$ & $0.642$ & rejected \\
$551$ & $0.604$ & $2.19\times$ & $0.161$ & $0.571$ & rejected \\
$74$ & $0.582$ & $3.17\times$ & $0.085$ & $0.574$ & rejected \\
$115$ & $0.543$ & $8.53\times$ & $0.007$ & $0.539$ & rejected \\
$289$ & $0.516$ & $3.90\times$ & $0.003$ & $0.512$ & rejected \\
\hline
\end{tabular}
\caption{Discovery-split rankings for the clean and auxiliary-dominated contrasts. AUROC measures event-level discrimination, top-$1\%$ enrichment is the prevalence of the positive class in the highest-activation tail relative to its overall prevalence, firing rate is $P(z_j>0)$, and ``Min. ctrl AUROC'' is the lowest AUROC across the matched-control schemes. Status records the role assigned after applying the fixed discovery criteria.}
\label{tab:app-readout-isolation}
\end{table}

\subsection{Latent-family definitions}
\label{app:latent-families}
\begin{table}[h]
\centering
\small
\begin{tabular}{l l r}
\hline
Family & Definition & $n$ \\
\hline
\texttt{clean\_pair} & $\{640,\,1006\}$ & $2$ \\
\texttt{clean\_family} & All $\mathrm{ams}\leq-0.85$, excluding the brightness-support coordinate $\zbs$ & $194$ \\
\texttt{clean\_core} & Members of \texttt{clean\_family} with firing rate $\geq0.05$ & $5$ \\
\texttt{aux\_family} & All $\mathrm{ams}\geq+0.85$ & $135$ \\
\hline
\end{tabular}
\caption{Reference-dictionary families. The final column gives the number of included latents.}
\label{tab:app-families}
\end{table}
The family interventions test whether a single-latent null could arise because a concept is distributed across several coordinates with similar response profiles. Table~\ref{tab:app-families} gives the reference-dictionary definitions used in these tests.

\subsection{PCA comparison}
\label{app:pca}
Table~\ref{tab:pca-sae-comparison} compares PC10, $\zbc$, $\zaux$, and $\zbs$ on the same validation events. For clean events the SAE's sparse latent $\zbc$ is a sharper discriminant than PC10, while for auxiliary-dominated events, $\zaux$ and PC10 perform about the same. For high-charge events, PC10 clearly outperforms $\zbs$ because $\zbs$ tracks charge only in its population-level average and is a poor per-event classifier, which is exactly why we do not treat it as a validated brightness read-out.

\begin{table}[h]
\centering
\begin{tabular}{llcc}
\hline
Positive class & Score & Test AUROC & Control AUROC \\
\hline
clean & PC$10$ & $0.846$ & $0.727$ \\
clean & $\zbc$ & $0.911$ & $0.787$ \\
auxiliary & PC$10$ & $0.847$ & $0.845$ \\
auxiliary & $\zaux$ & $0.840$ & $0.854$ \\
high charge & PC$10$ & $0.848$\tablefootnote{High/low charge is a binary contrast between the top and bottom quintiles of total charge.} & $0.750$ \\
high charge & $\zbs$ ($973$) & $0.532$ \tablefootnote{$\zbs$ is strongly charge-monotonic in population means but has near-chance per-event high/low-charge discrimination. This is why we describe it as brightness-associated rather than as a validated brightness read-out.} & $0.516$ \\
\hline
\end{tabular}
\caption{PCA and SAE scores evaluated on the independent \texttt{concept\_test} split. Control AUROC is the strictest matched-control value for each target.}
\label{tab:pca-sae-comparison}
\end{table}

\subsection{Distributed and exploratory physical information}

\subsubsection{Morphology proxies}
\label{app:morphology-probes}

To test whether linearly decodable morphology information is concentrated in a few SAE coordinates, we fit $k$-sparse probes to time extent, linefit speed, linearity, charge concentration, and vertical extent. A LARS path fitted on \texttt{concept\_dev} selects the $k$ coordinates, after which a ridge model is refitted on those columns and evaluated on \texttt{concept\_test}. Dense ridge probes on the raw \texttt{CLS} state and on all SAE coordinates provide reference ceilings.

\begin{table}[h]
\centering
\small
\begin{tabular}{lcccc|cccc}
\hline
 & \multicolumn{4}{c|}{Layer 8} & \multicolumn{4}{c}{Layer 3} \\
Target & \texttt{CLS} & SAE & $k{=}1$ & $k_{90\%}$ & \texttt{CLS} & SAE & $k{=}1$ & $k_{90\%}$ \\
\hline
linefit speed        & $0.48$ & $0.42$ & $0.16$ & $128$ & $0.61$ & $0.53$ & $0.14$ & $64$ \\
linearity            & $0.52$ & $0.53$ & $0.12$ & $128$ & $0.49$ & $0.52$ & $0.10$ & $128$ \\
time extent          & $0.78$ & $0.78$ & $0.36$ & $32$  & $0.88$ & $0.88$ & $0.43$ & $16$ \\
top-5 charge frac.   & $0.60$ & $0.51$ & $0.05$ & $256$ & $0.73$ & $0.64$ & $0.08$ & $64$ \\
$z$ RMS              & $0.67$ & $0.67$ & $0.10$ & $256$ & $0.74$ & $0.71$ & $0.19$ & $64$ \\
\hline
\end{tabular}
\caption{Held-out morphology-probe results at layers~8 and~3. Within each layer, \texttt{CLS} and SAE are the $R^2$ values of dense ridge probes using the raw \texttt{CLS} representation and all SAE latents, respectively. The $k{=}1$ column gives the best single-latent $R^2$, while $k_{90\%}$ is the smallest number of selected latents needed to reach $90\%$ of the dense-SAE $R^2$.}
\label{tab:sparse-morphology}
\end{table}
Table~\ref{tab:sparse-morphology} shows that time extent is moderately sparse, whereas the track-likeness and charge-concentration proxies require tens to hundreds of coordinates. The selected $k$ values are upper bounds rather than exact circuit sizes because the selection path is greedy. Thus, the absence of a validated single morphology latent does not imply that morphology information is absent from the representation, however it is mostly distributed across the dictionary.

\subsubsection{Layer-dependent depth read-outs}

\begin{table}[h]
\centering
\small
\begin{tabular}{c c c c c}
\hline
Layer & Latent & Depth AUROC & Min. ctrl AUROC & Direction-head $\Delta_+$ \\
\hline
$2$ & $171$ & $0.986$ & $0.984$ & $+0.0064^\circ$ \\
$3$ & $191$ & $0.925$ & $0.899$ & $-0.009^\circ$ \\
$8$ & \emph{none retained} & $0.598$ & --- & --- \\
\hline
\end{tabular}
\caption{Depth isolation across layers. ``Min. ctrl AUROC'' is the lowest AUROC across the matched-control schemes, and $\Delta_+$ is the direction-head angular-error shift after removing the selected latent. At layer~8, $0.598$ is the best raw AUROC, but no latent passed the isolation criteria.}
\label{tab:app-depth-layers}
\end{table}
The final-layer dictionary contains no validated depth coordinate (Table~\ref{tab:app-depth-layers}), but this depends on the representational site. Repeating the same isolation procedure at layers~2 and~3 recovers strong depth read-outs, with the layer-2 latent reaching AUROC $0.986$ and remaining essentially unchanged under matched controls. Neither early-layer coordinate passes the direction-head causal criteria. Geometry is therefore sharply localized at an earlier \texttt{CLS} site without becoming a direction-head bottleneck.

\subsubsection{Exploratory geometry profiles} \label{app:exploratory-geometry}  

We also search for latents whose activation changes with event direction, without first selecting candidates using a binary concept contrast. For azimuth, we divide the horizontal charge-centroid direction $\phi_c$ into $24$ bins and compute the mean activation of each latent in every bin. We then determine whether the resulting profile has one broad preferred direction ($k=1$), two opposite preferred directions ($k=2$), or the sixfold pattern expected from IceCube's hexagonal string layout ($k=6$).  Of the $91$ latents active enough for this test, $61$ vary across azimuth with an anisotropy of at least $0.15$. Among them, $52$ prefer one broad direction and $9$ show a two-direction pattern. None shows the expected sixfold detector pattern. The observed azimuthal dependence is therefore more likely to reflect the non-uniform distribution of events in the sample than the hexagonal detector geometry.  We perform a similar scan in zenith by measuring how each latent's mean activation changes with $\cos\theta_z$. We find $207$ latents with a strong monotonic response, $|\rho_{\mathrm{Spearman}}|\geq0.85$. However, the strongest responses come from dense latents that contribute broadly to reconstruction, and no individual latent survives the matched-control validation test. These direction-dependent latents are therefore reported as exploratory candidates rather than validated physical read-outs.

\subsection{Robustness across dictionaries}
\label{app:cross-seed}

We repeat the read-out analysis across four independently trained dictionaries. Table~\ref{tab:app-crossseed} groups functional analogues by role rather than by latent index. The clean-primary role passes every control scheme in three of four draws. The auxiliary role is less stable: only the reference seed passes all evaluable controls, and seed~3 yields no isolated auxiliary candidate. In contrast, the dense, charge-monotonic, non-selective support phenotype appears in every draw.

\begin{table}[h]
\centering
\small
\begin{tabular}{l l c c c c}
\hline
Role & Quantity & Seed 1 & Seed 2 & Seed 3 & Seed 4 \\
\hline
Clean primary & latent & $640$ & $754$ & $617$ & $947$ \\
 & test AUROC & $0.911$ & $0.874$ & $0.912$ & $0.996$ \\
 & min. ctrl AUROC & $0.787$ & $0.740$ & $0.781$ & $0.987$ \\
 & all ctrl schemes & Yes & No & Yes & Yes \\
\hline
Clean secondary & latent & $1006$ & $662$ & $918$ & $808$ \\
 & test AUROC & $0.694$ & $0.744$ & $0.691$ & $0.679$ \\
 & min. ctrl AUROC & $0.837$ & $0.697$ & $0.836$ & $0.831$ \\
 & all ctrl schemes & Yes & No & Yes & Yes \\
\hline
Auxiliary & latent & $195$ & $539$ & \emph{none} & $706$ \\
 & test AUROC & $0.840$ & $0.748$ & --- & $0.753$ \\
 & min. ctrl AUROC & $0.854$ & $0.715$ & --- & $0.731$ \\
 & all ctrl schemes & Yes & No & --- & No \\
\hline
Brightness support & carrier & $973$ & $119$ & $661$ & $781$ \\
 & charge monotonicity & $0.915$ & $0.915$ & $0.891$ & $0.915$ \\
 & clean--aux AUROC & $0.562$ & $0.569$ & $0.545$ & $0.568$ \\
\hline
\end{tabular}
\caption{Functional read-out analogues across SAE seeds. Each block reports the coordinate and its held-out validation quantities.}
\label{tab:app-crossseed}
\end{table}

The larger response landscape is more stable than the individual coordinates. Every seed contains more than $100$ clean-monotonic and more than $100$ auxiliary-monotonic latents at $|\mathrm{ams}|\geq0.85$. Across the tested expansion and sparsity grid, every dictionary also contains a control-surviving bright-clean analogue, with validation AUROC between $0.84$ and $0.99$. The robust object is therefore the physical role and the broader response axis, not a particular latent index.

\section{Head-relative intervention details}
\label{app:causal-tests}

\subsection{Family-level removals}
\label{app:direction-removals}

Table~\ref{tab:app-family} reports family removals on \texttt{causal\_test}. None of the clean- or auxiliary-side sets produces a selective direction-head effect. The control means are much larger than the medians for several families because a minority of high-density control sets removes substantial reconstruction mass. 

\begin{table}[h]
\centering
\small
\begin{tabular}{l r c r r r r r r}
\hline
Family & $n$ & Status & $\Delta_+$ & $\Delta_-$ & Select. & $z_+$ & Ctrl med. & Ctrl mean \\
\hline
\texttt{clean\_pair} & $2$ & null & $+0.033^{\circ}$ & $-0.007^{\circ}$ & $+0.039^{\circ}$ & $-0.33$ & $+0.024^{\circ}$ & $+0.844^{\circ}$ \\
\texttt{clean\_family} & $194$ & null & $-0.020^{\circ}$ & $-0.016^{\circ}$ & $-0.005^{\circ}$ & $-1.25$ & $+7.651^{\circ}$ & $+9.239^{\circ}$ \\
$\zaux$ & $1$ & null & $+0.004^{\circ}$ & $+0.033^{\circ}$ & $-0.029^{\circ}$ & $-0.22$ & $-0.006^{\circ}$ & $+0.247^{\circ}$ \\
\texttt{aux\_family} & $135$ & null & $-0.008^{\circ}$ & $+0.031^{\circ}$ & $-0.039^{\circ}$ & $-0.75$ & $+0.025^{\circ}$ & $+1.305^{\circ}$ \\
\texttt{clean\_core} & $5$ & null & $+0.018^{\circ}$ & $-0.010^{\circ}$ & $+0.028^{\circ}$ & $-0.63$ & $+0.033^{\circ}$ & $+2.966^{\circ}$ \\
\hline
\end{tabular}
\caption{Reference-dictionary removals on \texttt{causal\_test}. ``Select.'' is $\Delta_+-\Delta_-$; the last columns summarize matched controls.}
\label{tab:app-family}
\end{table}

The clean-pair removal is the more reliable verdict for the secondary coordinate $1006$: removing the pair changes the positive-class error by only $+0.033^\circ$ and fails both the effect and control-normalized criteria. Removing the complete $194$-latent clean family is also null. The absence of an effect therefore does not result from testing only one sparse coordinate.

\subsection{Cross-seed single-latent replication}
\label{app:cross-seed-causality}

We identify the latent that plays the same physical role in each dictionary and repeat the direction-head removal test. Table~\ref{tab:app-crossseed-causal} shows that removing the primary clean or auxiliary read-out never passes the causal criteria in any evaluable seed. The direction-head null is therefore not specific to the reference dictionary.

\begin{table}[h]
\centering
\small
\begin{tabular}{l c c c c}
\hline
Role & Seed 1 & Seed 2 & Seed 3 & Seed 4 \\
\hline
Clean primary & $+0.034^{\circ}$ (null) & $+0.029^{\circ}$ (null) & $-0.039^{\circ}$ (null) & $-0.011^{\circ}$ (null) \\
Clean secondary & $+0.024^{\circ}$ (null) & $+0.044^{\circ}$ (null) & $-0.050^{\circ}$ (null) & $-0.021^{\circ}$ (null) \\
Auxiliary & $+0.004^{\circ}$ (null) & $+0.004^{\circ}$ (null) & --- & $-0.026^{\circ}$ (null) \\
Brightness support & $+13.51^\circ$ (V) & $+14.81^\circ$ (V) & $+11.61^\circ$ (V) & $+14.59^\circ$ (V) \\
\hline
\end{tabular}
\caption{Direction-head shifts across SAE seeds. Parentheses give the fixed-rule verdict or dose-response shape.}
\label{tab:app-crossseed-causal}
\end{table}

The brightness-support coordinates behave differently. Removing them changes the angular error by $11.6^\circ$-$14.8^\circ$ in every seed. Their full dose responses are V-shaped: moving the activation either below or above its natural value damages the prediction. These coordinates therefore support direction reconstruction, but the head does not interpret their activation as a monotonic brightness signal.

\subsection{Writing and swap interventions}
\label{app:direction-writing}

Removal tests whether the direction head uses a read-out. Writing tests whether changing that read-out can change the head's prediction. We write clean-event activations into auxiliary events, write auxiliary-event activations into clean events, and swap brightness-related activations between bright and dim clean events. Table~\ref{tab:app-patching} reports the results. None of these interventions produces a larger effect than the matched controls. We therefore find no evidence that the direction head uses or responds to these validated read-outs.

\begin{table}[h]
\centering
\small
\begin{tabular}{l r r r r c}
\hline
Experiment & $\Delta$ recip. & $\Delta$ donor & $z(\Delta_+)$ & $z(\mathrm{sel.})$ & Verdict \\
\hline
write $640$ into aux              & $-0.012^{\circ}$ & $+0.025^{\circ}$ & $-0.92$ & $-0.71$ & null \\
\quad full swap ($+$zero $195$)   & $-0.002^{\circ}$ & $+0.034^{\circ}$ & $-0.01$ & $-0.51$ & null \\
write $\{640,1006\}$ into aux     & $+0.003^{\circ}$ & $+0.045^{\circ}$ & $-0.21$ & $+0.03$ & null \\
\quad full swap                   & $+0.013^{\circ}$ & $+0.053^{\circ}$ & $+0.38$ & $-0.22$ & null \\
write clean core ($5$ lat.)       & $+0.006^{\circ}$ & $+0.044^{\circ}$ & $-0.06$ & $+0.25$ & null \\
\quad full swap                   & $+0.016^{\circ}$ & $+0.051^{\circ}$ & $+0.28$ & $+0.56$ & null \\
write $195$ into clean            & $+0.024^{\circ}$ & $-0.003^{\circ}$ & $-0.53$ & $-1.04$ & null \\
\quad full swap ($+$zero $640,1006$) & $+0.034^{\circ}$ & $-0.003^{\circ}$ & $+0.21$ & $+0.69$ & null \\
brightness swap: dim$\to$bright ($640$) & $+0.016^{\circ}$ & $+0.114^{\circ}$ & $-0.44$ & $-0.40$ & null \\
brightness swap: bright$\to$dim ($640$) & $+0.115^{\circ}$ & $+0.013^{\circ}$ & $+0.10$ & $+0.70$ & null \\
\hline
\end{tabular}
\caption{Direction-head writing tests. Recipient and donor columns give the induced angular-error shifts.}
\label{tab:app-patching}
\end{table}

\subsection{Functionally trained dictionaries}
\label{app:e2e-dictionaries}
For each dictionary, we first identify the latents whose removal changes the direction-head output beyond matched controls. We call these latents the causal set. We then test every latent in this set against the full concept vocabulary, including event quality, auxiliary activity, energy, direction, depth, morphology, and angular error. Separately, we search the complete dictionary for a latent representing bright clean events. Table~\ref{tab:app-e2e-map} summarizes the four dictionaries.
\begin{table}[h]
\centering
\small
\begin{tabular}{l l c c c c}
\hline
Dictionary & Objective & \begin{tabular}[b]{@{}c@{}}Bright-clean\\ AUROC\end{tabular} & \begin{tabular}[b]{@{}c@{}}Causal\\ set\end{tabular} & \begin{tabular}[b]{@{}c@{}}Best concept\\ AUROC\end{tabular} & \begin{tabular}[b]{@{}c@{}}Validated\\ pairs\end{tabular} \\
\hline
Pure A & pure functional & $0.53$ & $16$ & --- & $0$ \\
Pure B & pure functional & $0.53$ & $11$ & $0.59$ & $0$ \\
Anchored A & reconstruction anchored & $0.87$ & $6$ & $0.69$ & $0$ \\
Anchored B & reconstruction anchored & $0.878$ & $6$ & $<0.75$ & $0$ \\
\hline
\end{tabular}
\caption{Results for the functionally trained dictionaries. `Bright-clean AUROC'' gives the best bright-clean read-out, `Causal set'' counts latents that affect the direction head, and `Best concept AUROC'' tests those latents against the full concept vocabulary. `Validated pairs'' counts the pairs that pass the complete validation procedure.}

\label{tab:app-e2e-map}
\end{table}
The two pure functional dictionaries contain $16$ and $11$ causal latents, confirming that the training objective finds coordinates that affect the head. However, neither dictionary contains an identifiable bright-clean latent: the best AUROC is $0.53$, close to chance. Adding the reconstruction term recovers a bright-clean latent in both anchored dictionaries, with AUROC $0.87$ and $0.878$. In both cases, this latent lies outside the causal set and its removal does not affect the direction head. Conversely, none of the causal latents passes validation for any tested physical concept. The result therefore persists even when causal latents and an identifiable bright-clean latent coexist in the same dictionary.

\subsection{Uncertainty-head replication}
\label{app:uhead-crossseed}

The reference-dictionary results are summarized in Table~\ref{tab:two-heads}. Here we test whether the causal effect depends on the SAE draw. Removal of $\zbc$, the clean pair, the clean family, and the clean core shifts the uncertainty head toward larger predicted error in every seed. The auxiliary family lowers the predicted error in every seed ($-0.085$ to $-0.093$), while single-latent $\zaux$ keeps its sign but falls below the effect threshold outside the reference dictionary.

\begin{table}[h]
\centering
\small
\begin{tabular}{lcccc}
\hline
Intervention & Seed 1 & Seed 2 & Seed 3 & Seed 4 \\
\hline
remove $\zbc$ & $+0.310$ & $+0.193$ & $+0.295$ & $+0.612$ \\
remove clean pair & $+0.321$ & $+0.374$ & $+0.307$ & $+0.612$ \\
remove clean family & $+0.775$ & $+0.837$ & $+1.012$ & $+0.805$ \\
remove clean core & $+0.373$ & $+0.493$ & $+0.681$ & $+0.617$ \\
clean core after support exclusion & $+0.373$ & $+0.493$ & $+0.302$ & $+0.617$ \\
remove auxiliary family & $-0.085$ &$-0.089$ & $-0.093$ & $-0.091$ \\
remove $\zaux$ & $-0.059$ &$-0.011$ & -- & $-0.015$ \\

\hline
\end{tabular}
\caption{Cross-seed uncertainty-head removals. Entries are shifts $M_1$ in predicted log error.}
\label{tab:app-uhead-crossseed}
\end{table}

The first four rows reproduce the positive-sign reference-seed verdicts in every draw. The fifth row diagnoses the larger seed-3 core effect: after removing its support-carrier contamination, $M_1$ falls from $+0.681$ to $+0.302$ but remains significant ($z=6.8$). Seed~3 also splits the support role between latents $661$ and $144$. The direction head is more sensitive to $661$, whereas the uncertainty head is more sensitive to $144$. Consequently, the pre-specified single-carrier uncertainty dose replicates in $3/4$ seeds; a post-hoc joint dose on $\{661,144\}$ restores the monotone response in the remaining draw.
\stopappendices
\end{document}